\documentclass[useAMS,usenatbib]{mn2e}
\newcommand{\dev}{\ensuremath{\,\mathrm{d}}}
\usepackage{amssymb}
\usepackage{amsmath}
\usepackage[dvips]{graphicx}
\usepackage{epsfig}
\usepackage{multicol}

\title{Self-similar relativistic blast waves with energy injection}

\author[H.J. van Eerten]{Hendrik van Eerten$^1$\thanks{Alexander von Humboldt Fellow} \thanks{e-mail: hveerten@mpe.mpg.de} \\
$^{1}$Max-Planck-Institut f\"ur Extraterrestrische Physik, Giessenbachstra\ss e 1, 85748 Garching, Germany}

\begin{document}

%\date{}

\pagerange{\pageref{firstpage}--\pageref{lastpage}}

\maketitle

\label{firstpage}

\begin{abstract}
A sufficiently powerful astrophysical source with power law luminosity in time will give rise to a self-similar relativistic blast wave with a reverse shock traveling into the ejecta and a forward shock moving into the surrounding medium. Once energy injection ceases and the last energy is delivered to the shock front, the blast wave will transit into another self-similar stage depending only on the total amount of energy injected.

I describe the effect of limited duration energy injection into environments with density depending on radius as a power law, emphasizing optical / X-ray Gamma-ray Burst afterglows as applications. The blast wave during injection is treated analytically, the transition following last energy injection with one-dimensional simulations. Flux equations for synchrotron emission from the forward and reverse shock regions are provided. The reverse shock emission can easily dominate, especially with different magnetizations for both regions. Reverse shock emission is shown to support both the reported X-ray and optical correlations between afterglow plateau duration and end time flux, independently of the luminosity power law slope. The model is demonstrated by application to bursts 120521A and 090515, and can accommodate their steep post-plateau light curve slopes.
\end{abstract}


\section{Introduction}
\label{introduction_section}

Some cataclysmic astrophysical events, such as the merger of neutron stars \citep{Eichler1989, Paczynski1991} or the collapse of a very massive star \citep{Woosley1993, Paczynski1998, MacFadyen1999}, can give rise to brief flashes of gamma rays (`gamma-ray bursts', or GRBs) that can be detected at cosmological distances. A common feature of GRB models is the launch of collimated relativistic ejecta that interact with the surrounding medium and produce an afterglow signal that can be detected from X-rays to radio as the blast wave decelerates. 

The classical fireball model for the afterglow (e.g. \citealt{Meszaros1993, Meszaros1997}) describes the relativistic hydrodynamical evolution after a mass and energy ($10^{48-51}$ erg, depending on the progenitor type) are injected effectively instantaneously into a small region. The fireball expands, accelerates and ultimately decelerates, by sweeping up external matter, with a self-similar relativistic fluid profile first described by \cite{Blandford1976} (hereafter denoted BM76). This deceleration phase might be preceded by a brief phase where a reverse shock runs into the original ejecta (see e.g. \citealt{Sari1995, Sari1997, Kobayashi1999}). The main observed emission component is synchrotron radiation from shock-accelerated electrons interacting with small-scale shock-generated magnetic fields, giving rise to a broadband signal that follows a power-law decay (or rise, at low observer frequencies) in time (e.g. \citealt{Blandford1977}). Standard GRB afterglow theory has recently been reviewed by various authors, including \cite{Piran2004, Meszaros2006, Granot2007, vanEerten2013proceedings}.

Since the launch of \emph{Swift} \citep{Gehrels2004}, early time plateau phases of shallow decay in afterglow X-ray light curves have been revealed to be far more common than originally expected. These plateaus are commonly attributed to prolonged injection of energy (see e.g. \citealt{Nousek2006, ZhangBing2006}), which can take different forms, such as ejecta with a range of velocities catching up with the shock front (e.g. \citealt{ReesMeszaros1998, Panaitescu1998, SariMeszaros2000}), long-term luminosity of the source (e.g. \citealt{ZhangMeszaros2001}) or conversion of Poynting flux from the ejecta (e.g. \citealt{Usov1992, Thompson1994, MeszarosRees1997Poynting, LyutikovBlandford2003}). When the blast wave is continously driven from the back, the impulsive energy injection scenario no longer correctly describes the fluid evolution and the blast wave will decelerate more slowly or not at all. Instead, for the duration of injection of energy, a more complex system of shocks will form similar to the brief reverse shock stage for massive ejecta, with a contact discontinuity separating the ejecta from the swept-up ambient medium and a reverse shock running into the ejecta, in addition to the forward shock running into the medium that is also present in the impulsive injection scenario.

The exact location and nature of emission during the plateau phase is still not fully resolved. The observed emission will be a mixture of forward and reverse shock contributions, and either can be dominant (see e.g. \citealt{ZhangMeszaros2001, ZhangBing2006, Nousek2006, ButlerKocevski2007}). Treatments of outflows with a long-lived reverse shock have shown that it is possible to account for a significant part of the overall observed flux and light curve features by emission from the reverse shock region \citep{Uhm2011, Uhm2012, Leventis2014}. Another strong indication that forward shock emission alone is insufficient to explain observations is the abrupt drop in luminosity that is sometimes seen at the end of the plateau phase (e.g. \citealt{Troja2007, Rowlinson2013}).

In this study I consider the dynamics, evolution and emission from energy injection by a power law luminosity from the central source, which includes and generalizes the `thick-shell' case for massive ejecta. In the case of a sufficiently long-lived reverse shock that has become relativistic in the frame of the inflowing fluid, the full reverse shock / forward shock profile is self-similar and treated generally by BM76 (who omit only the reverse shock region density profile). Separate self-similar solutions exist in the literature (e.g. 
\citealt{Nakamura2006, Nakayama2005}), but for our purpose, the radial flow with relativistic reverse shock described in BM76 is sufficient. Since the injection of energy still lasts only for a limited time, the transition from sustained to effectively impulsive injection (i.e. when the injection timescale becomes negligible again compared to the explosion duration) is discussed in detail, including a numerical hydrodynamics approach. Flux equations are derived for the emission that show the relative contributions from forward and reverse shock regions. These equations can be applied directly to observational X-ray and optical data. The reverse shock emission is found to often be important, as is demonstrated using `typical' long GRB afterglow parameters and short GRBs 090515 and 120521A.

In section \ref{self-similar_section}, the self-similar fluid profile during energy injection from a power-law luminosity source is derived and placed in the context of the fireball model. In section \ref{transition_section}, the transition after cessation of energy injection is discussed both analytically and numerically (using the \textsc{ram} relativistic hydrodynamics code, \citealt{ZhangWeiqun2006}). Flux equations and the role of the reverse shock region for the observed emission are treated in section \ref{light_curve_section}, including an application to GRBs 090515 and 120521A. A general discussion follows in section \ref{discussion_section}, a summary in section \ref{summary_section}, and some technicalities are deferred to appendices.

In a separate study, \cite{Leventis2014} also find an important role for the reverse shock emission in shaping the observed flux. They consider a homogeneous shell model for long-lived reverse shocks and assume a homogeneous circumburst medium, whereas this study consideres a generic power law medium, a full self-similar fluid profile and also examines the transition following energy injection using numerical simulations. Earlier work by \cite{Uhm2012} considers the case where energy injection results from shells with a range of Lorentz factors catching up with the forward shock. There, the fluid profile is not self-similar but calculated in detail using the `mechanical model' approach to afterglow blast waves from \cite{Beloborodov2006}.

\section{Self-similar solution for relativistic blast wave with energy injection}
\label{self-similar_section}

In this section, I discuss the dynamics of purely radial flow during energy injection.

\begin{figure}
 \centering
  \includegraphics[width=\columnwidth]{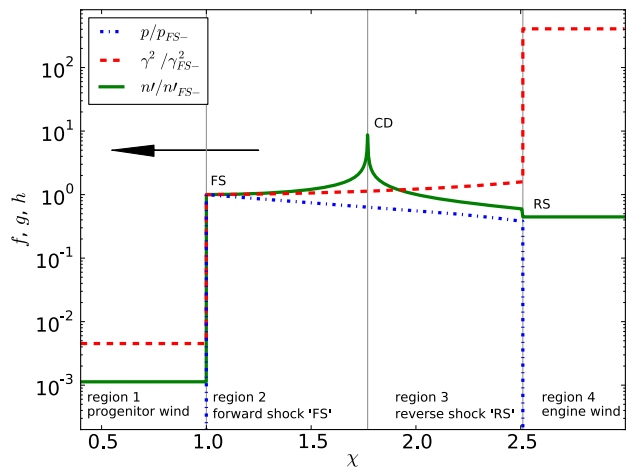}
\caption{Self-similar fluid profile for a blast wave in a stellar wind environment ($k=2$) and continuous energy injection $q=0$. In region 2, the profiles as shown above are equivalent to self-similar functions $f$, $g$, $h$. In region 3, $f$, $g$ are still equivalent to what is plotted, while the density profile differs from self-similar function $H$ by a normalization factor. At fixed time $t$ we have $\textrm{d} r / ct = - \textrm{d} \chi / 2 \Gamma^2$. The direction of the blast wave is to the left. The relative values outside of the shocked region and across the contact discontinuity (for the density) are set according to the typical values discussed in section \ref{typical_values_section}. In a stellar wind environment, the fluid profile as plotted here remains completely unchanged over time.}
 \label{windprofile_figure}
\end{figure}

\subsection{Free-flowing relativistic wind}
\label{free_flowing_wind_subsection}

I take as starting point a prolonged injection of energy at time $t_{in}$ at initial radius $R_0$, according to
\begin{equation}
L = L_0 t_{in}^q.
\label{luminosity_equation}
\end{equation}
At the same time matter is injected according to
\begin{equation}
L_M = L_{M,0} t_{in}^{-s}.
\label{matter_inflow_equation}
\end{equation}
In case of a collimated outflow, these injection rates refer to \emph{isotropic equivalent} luminosities, since at early stages the flow is purely radial.
These generalize the impulsive injection of the fireball scenario, for which $L \to L \delta (t-t_0)$ and $L_M \to L_M \delta (t-t_0)$. 
When the time evolutions of mass and energy injections are the same, the ratio between $L$ and $L_M$ is given by a $\eta \equiv L / L_M c^2$, again generalizing $\eta \equiv E / M c^2$ for the fireball, where $E$ the total explosion energy, $M$ the total explosion mass and $c$ the speed of light. I will consider only cases where the baryon loading is small, and $\eta \gg 1$. Throughout the paper I will assume $q = -s$ for simplicity. If $q \neq -s$, but $\eta \gg 1$ is maintained throughout the injection, the argument remains essentially unchanged, albeit that an extra factor $t^s$ needs to be accounted for in the RS density profile and in the flux equations. For sources such as GRB's, the true relative time evolution of energy and matter injection is not known.

At small radii this results in an accelerating relativistic wind profile, following the dynamics of the small Lorentz-contracted shell of the impulsive energy injection fireball \citep{Piran1993, Kobayashi1999} throughout its radial profile. At a radius $R_L \equiv \eta R_0$, all internal energy in the outflow is converted into kinetic energy (for the fireball shell analog, see \citealt{Goodman1986, Paczynski1986, Shemi1990, Kobayashi1999}) and the outflow will eventually proceed with fixed Lorentz factor according to:
\begin{equation}
 \gamma = \eta, \; \rho = \frac{L_M}{4 \pi c \eta r^2}, \; \; p = \frac{L R_0^{2/3}}{16 \pi c \eta^{4/3} r^{8/3}}.
\label{freely_expanding_flow_equations}
\end{equation}

These equations describe the fluid profile for a freely expanding outflow in a dilute medium. However, in reality the expansion does not occur in a total vacuum and at the surface of the sphere a termination shock profile is formed consisting of a reverse shock (RS) moving into the freely expanding ejecta (but still moving outward in the `lab' frame centered on the origin of the explosion), a contact discontinuity (CD) separating the wind from the environment and a forward shock (FS) moving into the circumburst environment, likely a stellar wind profile shaped by the progenitor system or a homogeneous interstellar medium (ISM) type profile.

\subsection{Self-similar blast wave with energy injection}
\label{self-similar_blast_wave_subsection}

A self-similar profile for the RS-CD-FS system is provided by \cite{Blandford1976} for the case of a relativistic FS and relativistic RS. In BM76 (I will use `BM76' to refer to the paper, `BM solution' to refer to the self-similar solution), the density profile inside of the contact discontinuity is not discussed, and indeed most information about the flow can be derived assuming only eq. \ref{luminosity_equation}, while remaining agnostic about the mass flux (eq. \ref{matter_inflow_equation}). In BM76, the similarity variable is given by
\begin{equation}
 \chi = [1 + 2(m+1) \Gamma^2](1 - r / tc),
\label{chi_equation}
\end{equation}
increasing from 1 at the shock front to higher values downstream, for a blast wave with a forward shock Lorentz factor $\Gamma$ evolving in time $t$ according to $\Gamma^2 \propto t^{-m}$. 

Combining the shock-jump conditions for a strong relativistic forward shock with the assumption of self-similarity one obtains
\begin{eqnarray}
 p & = & \frac{2}{3} \omega_{FS+} \Gamma^2 f(\chi), \nonumber \\
 \gamma^2 & = & \frac{1}{2} \Gamma^2 g(\chi), \nonumber \\ 
 n' & = & 2 n_{FS+} \Gamma^2 h(\chi),
\label{BM_jump_equations}
\end{eqnarray}
between the FS and the CD, where the self-similar functions obey $f(1) = g(1) = h(1) = 1$ at the shock front\footnote{The subscript `$FS+$' implies that the a quantity is evaluated at radius $r \downarrow r_{FS}$, i.e. approaching the FS radius from above. We will likewise use `$FS-$' (approach from below) and `$FS$' (exactly at) for FS, CD and RS.}. For a cold medium, enthalpy $\omega_{FS+} = \rho_{FS+} c^2$.
The prime on the number density $n'$ refers to the frame at which the origin of the explosion is at rest (i.e. the frame of the circumburst medium or `lab' frame) and is related to the comoving number density $n$ according to $n = \gamma^{-1} n'$.
Mass and number density are related via $\rho = m_p n$ where $m_p$ the proton mass.
The self-similar functions can be expressed as differential equations when combining eqs. \ref{BM_jump_equations} with the equations of relativistic fluid dynamics.

If we now \emph{assume} $\eta$ to be constant in time \emph{and} consider only radii $r > R_L$, we can extend the analytical solutions for the differential equations from BM76 to include the density profile in the RS region. In terms of $x \equiv g \chi$, the solution to the density profile equation, in the case of ongoing injection, can be found to be
\begin{equation}
h(x) = C \times A(x)^{-\gamma_2} \times B(x)^{\mu_1 / \gamma_1} \times (2 - x)^{-\mu_2}.
\label{h_solution_equation}
\end{equation}
Here $A(x)$ and $B(x)$ are functions of $x$ obeying $A(1) = B(1) = 1$, while $\gamma_1$, $\gamma_2$, $\mu_1$ and $\mu_2$ are determined by the power law slope $k$ of the surrounding medium density profile (using $\rho \equiv \rho_{ref} (r/r_{ref})^{-k}$) and $q$. For completeness, these symbols and the analytical solution for the full set of fluid quantities are defined in appendix \ref{additional_symbols_appendix}. $C$ is a constant of integration whose value is determined by the boundary conditions. In the FS region, $C = 1$, from $h(1) = 1$. For given $q, k$, the positions of the CD and RS are fixed in self-similar coordinates and can be shown to be given by $x_{CD} = g(\chi_{CD}) \chi_{CD} = 2$ and $x_{RS} = g(\chi_{RS}) \chi_{RS} = 4$. The former follows from the constraint that all fluid elements outside of the CD have to originate at the shock front, the latter from the assumption of a relativistic reverse shock (in the frame of the inflowing wind).

However, at the CD the RS region is disconnected from the forward region. Extending the solution from BM76, we define a second function $H(x)$ to describe the self-similar profile in the RS region:
\begin{equation}
H(x) = C_1 \times A(x)^{-\gamma_2} \times B(x)^{\mu_1 / \gamma_1} \times (x - 2)^{-\mu_2},
\label{H_solution_equation}
\end{equation}
which differs from $h(x)$ by a constant factor that is chosen such that $H_{RS+} \equiv 1$. In some cases (e.g. the wind case with $q = 0$, $k = 0$, $m = 1$), the density profile singularity at the CD has density going up to infinity. This represents a breakdown of the underlying assumption $p \gg \rho c^2$, meaning that the self-similar solution therefore already ceases to be valid in the vicinity of this point and not just at the singularity itself.

The forward shock Lorentz factor can be shown to be
\begin{equation}
\Gamma^2 = \left[ \frac{L_0 \chi_{RS}^{1+q} c^{k-5}}{2^q  (m+1)^q 16 \pi \rho_{ref} r_{ref}^k f_{RS}} \right]^{\frac{1}{2+q}} t^{\frac{q+k-2}{2+q}},
\label{Gamma_equation}
\end{equation}
by equating the energy influx through the RS to the total energy in the RS and FS region. This also fixes $m$.
Energy and matter injected at time $t_{in}$ reach the reverse shock at radius $R_{RS}$ at time 
\begin{equation}
t = R_{RS} / c + t_{in},
\label{time_to_reach_equation}
\end{equation}
leading to
\begin{equation}
t = t_{in} 2 (m+1) \Gamma^2 \chi_{RS}^{-1}.
\label{t_and_t_in_equation}
\end{equation}

For the reverse shock we have $\Gamma_{RS} = \Gamma / \sqrt{\chi_{RS}}$ and
\begin{equation}
 \bar{\Gamma}_{RS} = \frac{1}{2} \left( \frac{ \Gamma \sqrt{\chi_{RS}}}{\gamma_{RS-}} + \frac{\gamma_{RS-}}{\Gamma \sqrt{\chi_{RS}}} \right),
\end{equation}
in the frame of the inflowing material at $r_{RS-}$ (i.e. directly ahead of the RS shock front), denoted with a bar.

The shock jump condition for the number density behind the (assumed relativistic) reverse shock is given by:
\begin{equation}
 \bar{n}_{RS+} = 2 n_{RS-} \bar{\Gamma}_{RS}^2,
\end{equation}
where $n_{RS-}$ the number density of the inflowing material before it crosses the reverse shock (comoving, so denoting it with a bar would be redundant). This can be expressed in the lab frame as:
\begin{equation}
n'_{RS+} = 2 n_{RS-} \gamma_{RS-}, 
\end{equation}
where $\gamma_{RS-}$ the Lorentz factor of the inflowing material in the lab frame. 
Here however we did use the assumption that $\gamma_{RS-} \gg \Gamma_{RS}$, a condition that is not met initially, when $R \ll R_0$, but easily met later on.

By construction, the density profile throughout the RS region is therefore given by
\begin{equation}
n' = 2 n_{RS-} \gamma_{RS-} H(x),
\end{equation}
where we maintain $x \equiv g \chi$ also throughout the RS region.

A quantity of interest is the ratio between densities behind the forward and reverse shock:
\begin{eqnarray}
\frac{n'_{RS+}}{n'_{FS-}} & = & \frac{\rho_{RS-} \gamma_{RS-}}{\Gamma^2 \rho_{FS+}} \nonumber \\
 & = & \frac{L_0}{4 \pi \rho_{ref} c^{5-k} R_{ref}^k \eta} \times t_{in}^q \times t^{k-2} \times \Gamma^{-2} \nonumber \\
 & = & \frac{1}{\eta} \left[ \frac{2^q c^{k-5} f_{RS}^{1+q} L_0}{\chi_{RS} \pi (m+1) \rho_{ref} R_{ref}^k} \right]^{\frac{1}{2+q}} t^{\frac{q+k-2}{2+q}}.
\label{density_ratio_equation}
\end{eqnarray}
using $R \sim ct$ in the second step. For continuous energy injection into a stellar wind environment  ($q=0$, $k=2$, $m=0$), this implies that the ratio between densities at the RS and FS stays constant in time. For a homogeneous circumburst environment ($q = 0$, $k = 0$, $m = 1$), this implies a decreasing density in the RS region 3 relative to FS region 2, with $n'_{RS+}/n'_{FS-} \propto t^{-1}$. Example fluid profiles for the wind and ISM case are shown in Figs. \ref{windprofile_figure} and \ref{ISM_profile_figure}.

\subsection{Typical values for GRB afterglows}
\label{typical_values_section}

\begin{figure}
 \centering
  \includegraphics[width=\columnwidth]{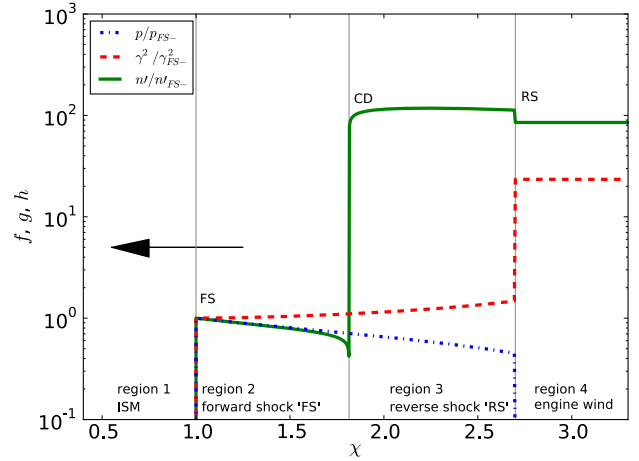}
\caption{Self-similar fluid profile for a blast wave expanding in a homogeneous medium ($k = 0$) and with constant energy injection ($q = 0$). The relative density difference between forward and reverse shock regions depends on time. For this figure we have used the ISM values described in section \ref{typical_values_section}, with time $t = 10^4$ seconds. The interpretation of the fluid profile in terms of self-similar functions $f$, $g$, $h$ and $H$ is the same as in Fig. \ref{windprofile_figure}.}
 \label{ISM_profile_figure}
\end{figure}

We can now plug in some values we expect to be representative of \emph{Swift} afterglows. If the total explosion energy in the blast wave $E_j = 10^{51}$ erg, for a pair of collimated blast waves that start out with collimation angle $\theta_0 = 0.1$ rad., and is injected over the course of $T_{in} = 10^4$ s. at a constant rate ($q = 0$), we have:
\begin{equation}
L = L_0 = E_{iso} / T_{in} = 2 E_j / (\theta_0^2 T_{in}) = 2\cdot 10^{49} \textrm{ erg s}^{-1}.
\end{equation}
Here $E_{iso}$ is the isotropic equivalent explosion energy, relevant for radial flow.
For a fireball starting at radius $R_0 = 10^{11}$ cm and with $\eta = 300$ (also its peak Lorentz factor), we have a coasting radius $R_L = 3 \cdot 10^{13}$ cm. For these values of $\eta$ and $L$ the mass loss rate $L_M = 1.18 \, M_\odot$ yr$^{-1}$, with a total mass los $M = 3.73 \times 10^{-4} \, M_{\odot}$ after $10^4$ s. Note that, while we consider 300 to be typical here, there is no unambiguous canonical value for $\eta$ that one can infer from the literature on GRB observations. There exists a range of methods for estimating the initial Lorentz factors of GRB ejecta (see e.g. \citealt{ZouPiran2010, Racusin2011} and references therein), but these are sensitive to underlying model assumptions about the prompt emission and the initial nature of the outflow.

For $q = 0$ and $k = 2$, we obtain\footnote{In table I of BM76 the corresponding entry for $K$, setting the square of the FS Lorentz factor, is wrong by a factor $\sqrt{3}$ and should be $K = 1.486$. The same $\sqrt{3}$ term is lacking in their eq. (71), but included in their eqs. (58), (59), (72).} $m = 0$, $\chi_{CD} = 1.77$, $\chi_{RS} = 2.51$, $f_{RS+} = 0.379$, $f_{CD} = 0.645$, $g_{RS+} = 1.59$ and $g_{CD} = 1.13$. The fluid profile in the wind case is shown in Fig. \ref{windprofile_figure}.

If we take $n_{ref} = 29.9$ cm$^{-3}$ and $R_{ref} = 10^{17}$ cm so that $\rho_{ref} R_{ref}^k = 5 \times 10^{11}$ g cm$^{-1}$, which follows for a progenitor stellar wind with velocity $10^3$ km s$^{-1}$ and mass loss of $10^{-5} M_\odot$ yr$^{-1}$ \citep{Chevalier2000, Granot2002}, we then obtain a constant forward shock Lorentz factor $\Gamma = 21.0$ and a RS Lorentz factor $\Gamma_{RS} = 13.3$.
In the frame of the inflowing wind (once $R_{RS} > R_L$), we have $\bar{\Gamma}_{RS} = 11.3$, which is indeed relativistic.

For a stellar wind environment the ratio of FS and RS densities determined by eq. \ref{density_ratio_equation} is constant and given by $n'_{RS+}/n'_{FS-} = 0.89$.

Using eq. \ref{t_and_t_in_equation} we find that the last of the injected energy at $T_{in}$ is delivered to the blast wave at $T = 3.52 \times 10^6$ s., at which point the outer radius of the explosion is $R = 1.05 \times 10^{17}$ cm.
This event is observed (ignoring redshift corrections) at $T_{obs} = T - R_{RS} / c$ (if we take $T_{obs} = 0$ to coincide with the time of the explosion. The expression otherwise merely states that emission departing closer to the observer is seen earlier than simultaneous emission from further distant). It therefore follows that $T_{obs} = T_{in}$ (cf. eq. \ref{time_to_reach_equation}), such that the observed duration of the plateau in \emph{Swift} data can be interpreted directly in terms of duration of energy injection (the argument is analogous to the link between observed variability from internal shocks in the prompt emission and internal engine variability).
From this point on the blast wave will evolve into an outflow described by the impulsive energy injection BM solution with energy $E_{iso}$.

If we take as a measure of the width $\Delta R$ of the impulsive energy blast wave the width of a homogeneous shell with density determined by the relativistic shock jump condition at the forward shock and total mass content equal to the total swept-up mass, we obtain $\Delta R = R / [ 2 (3-k) \Gamma^2 ]$ and $\chi_{B} = (7-2k) / (3-k)$ for the self-similar position of the back.

If the circumburst medium is homogeneous instead of a free-flowing stellar wind environment, the relative fluid densities of the FS and RS regions are time-dependent. If we use the same explosion energy and injection duration as above but set $n_{ref} = 1$ cm$^{-3}$, typical for values measured assuming a homogeneous interstellar medium (ISM) or stalled wind type environment with $k = 0$, we obtain the following.

The last energy injected at $T_{in} = 10^4$ s. is delivered to the blast wave at $T = 1.07 \times 10^7$ s. At this point the forward shock front is at radius $R = 3.20 \times 10^{17}$ cm. The FS has Lorentz factor $\Gamma = 26.8$, the RS has $\Gamma_{RS} = 16.3$, $\bar{\Gamma}_{RS} = 9.21$. Auxiliary quantities have the values $m = 1$,  $\chi_{CD} = 1.81$, $\chi_{RS} = 2.70$, $f_{RS+} = 0.449$, $f_{CD} = 0.709$, $g_{RS+} = 1.48$ and $g_{CD} = 1.10$. At time $T$, the density ratio between FS and RS is equal to $n'_{RS+}/n'_{FS-} = 0.160$. The ISM fluid profile (at an earlier time than $T$) is shown in Fig. \ref{ISM_profile_figure}.

\section{Transition after energy injection}
\label{transition_section}

After the last energy injected at $T_{in}$ has crossed the reverse shock, the fluid is expected to evolve from the self-similar energy injection profile to the instantaneous energy injection profile at $t \gg T_{in}$. Using the sound crossing time according to the BM solution as an estimate for the duration, the resulting transition time should be fairly quick. The comoving speed of sound is $c_s \equiv 1 / \sqrt{3}$ in the relativistic limit, and in the lab frame we have
\begin{equation}
 c'_s \approx 1 - \frac{1 - c_s}{1 + c_s} \frac{1}{g(\chi)\Gamma^2} \equiv 1 - \frac{\alpha_s}{g(\chi) \Gamma^2}.
\end{equation}
Using this value for $\dev r / \dev t$ to rewrite $\dev \chi / \dev t$, we arrive at
\begin{equation}
 \frac{\dev x}{(1+xQ(x))(2 \alpha_s - x)} = (m+1) \frac{\dev t}{t},
 \label{soundwave_equation}
\end{equation}
with $Q(x)$ defined by 
\begin{equation}
Q(x) \equiv \frac{(7m + 3k -4) - (m+2)x}{(m+1)(4-8x+x^2)}.
\end{equation}
In the impulsive energy case, where $g(\chi) = 1 / \chi$, the arrival time $t_{stop}$ at the front ($\chi = 1$), for sound waves departing from $\chi_{start}$ at $t_{start}$, is then given by
\begin{equation}
 t_{stop} = \chi_{start}^{\frac{-1}{(1+m)(2 \alpha_s - 1)}} t_{start}.
 \label{impulsive_crossing_time_equation}
\end{equation}
A sound wave departing from the `back' of the shock $\chi_B$ will arrive at $1.35 t_{start}$ when $k=0$, and at $1.55 t_{start}$ when $k = 2$. The analytical expression in the energy injection case is less clean, but some results are shown in table \ref{soundwave_table} for sound waves departing from the RS at $x \equiv 4$ and the CD at $x \equiv 2$. Note also that in the above, the limit $\alpha_s \downarrow 0$ corresponds to replacing $c_s$ by the speed of light, yielding $t_{stop} = \chi_{start}^{1/(m+1)} t_{start}$, for both impulsive and sustained energy injection profiles. This is the absolute minimum amount of time that the front of the shock will remain unaffected by changing conditions at the back. The other limit $a_s \uparrow 1$ corresponds to advective motion away from the shock front(s).

\begin{table}
\begin{center}
\begin{tabular}{|r|l|l|l|}
\hline
 & RS $\to$ FS & RS $\to$ CD & CD $\to$ FS \\
\hline
$k = 2$, $q = -1/2$ & 5.05 & 1.57 & 3.22 \\
$k = 2$, $q = 0$ & 4.05 & 1.55 & 2.61 \\
$k = 2$, $q = 1/2$ & 3.69 & 1.54 & 2.39 \\
\hline
$k = 0$, $q = -1/2$ & 2.38 & 1.29 & 1.84 \\
$k = 0$, $q = 0$ & 2.11 & 1.28 & 1.65 \\
$k = 0$, $q = 1/2$ & 2.01 & 1.28 & 1.57 \\
\hline
\end{tabular}
\caption{Arrival time factors $X$ for arrival at the shock front or CD for sound waves departing from RS or CD. The arrival times themselves are then given by $t_{stop} = X t_{start}$. The factors in the RS $\to$ FS column also follow from multiplying the values in the RS $\to$ CD and CD $\to$ FS columns.}
\label{soundwave_table}
\end{center}
\end{table} 

Once the cessation of energy injection has been communicated to the front of the shock, it is expected that the further evolution of the blast wave will start to resemble the impulsive energy injection scenario. The most important characteristics of the blast wave are its Lorentz factor and radius (also in terms of its observational signature, since a homogeneous shell approximation yields the correct temporal behavior and a flux level that differs from a more detailed approach by a constant factor). When all energy is injected, the shock Lorentz factor will eventually evolve according to (BM76):
\begin{equation}
\Gamma^2 = \frac{(17-4k) E_{iso}}{8 \pi \rho_{ref} c^{5-k} R_{ref}^k} t^{k-3},
\label{Gamma_impulsive_equation}
\end{equation}
which should be compared to eq. \ref{Gamma_equation}. The ratio between the two Lorentz factors, $\Gamma_{FS}$ from eq. \ref{Gamma_equation} and $\Gamma_{I}$ from eq. \ref{Gamma_impulsive_equation}, at time $t = X T_{last}$ (obtaining $T_{last}$ using eq. \ref{t_and_t_in_equation} and $X$ from table \ref{soundwave_table}), is found to be
\begin{equation}
\frac{\Gamma_{FS}^2}{\Gamma_{I}^2} = \frac{(m+1)(q+1)X^{q+1}}{(17-4k)f_{RS+}},
\end{equation}
which is independent of injection duration, explosion energy and circumburst structure. For $k=2$, $q=0$, we find $\Gamma_{FS}^2 / \Gamma_I^2 \approx 1.2$ and for $k=0$, $q=0$ we find $\Gamma_{FS}^2 / \Gamma_I^2 \approx 0.55$. It follows that at the time the sound wave reaches the front, the Lorentz factors for the two asymptotic regimes are already comparable. Based on that, no sudden jump or drop in fluid Lorentz factor is expected as the blast wave transits from one regime to the other.

\subsection{Transition Simulations in one dimension}
\label{transition_simulations_subsection}

Using the numerical approach described in appendix \ref{code_appendix}, a number of relativistic hydrodynamics (RHD) simulations have been run in one dimension of $q = 0$ energy injection into either a wind-shaped or homogeneous environment. In order to test the analytical predictions from the preceding section, four scenario's were explored, corresponding to different combinations of $k = 0$, $2$ and $T_{in} = 10^4$ s., $\infty$.

\begin{figure}
 \centering
  \includegraphics[width=\columnwidth]{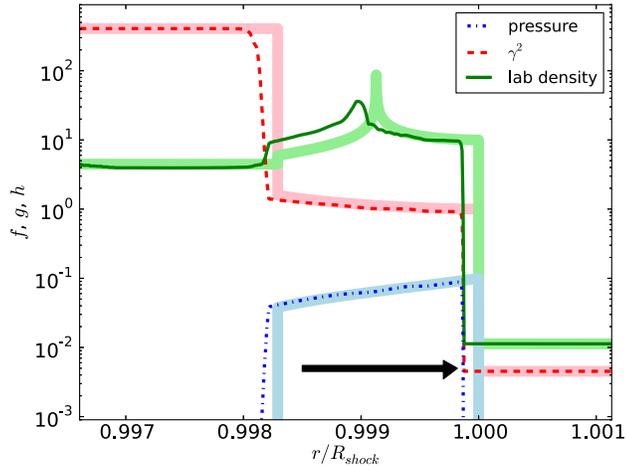}
\caption{Stellar wind profile for sustained injection scenario ($T_{in} = \infty$), at the time the last energy would have been delivered if $T_{in}$ had been $10^4$ s. The direction of the blast wave is to the right. pressure profile $p$, Lorentz factor-squared $\gamma^2$ and lab frame density have been scaled to $10^{-1}$, 1, $10$ at the FS, respectively. The radius had been scaled to the analytically expected FS radius. Thick light colored lines indicate the analytical solution.}
 \label{wind_before_figure}
\end{figure}

\begin{figure}
 \centering
  \includegraphics[width=\columnwidth]{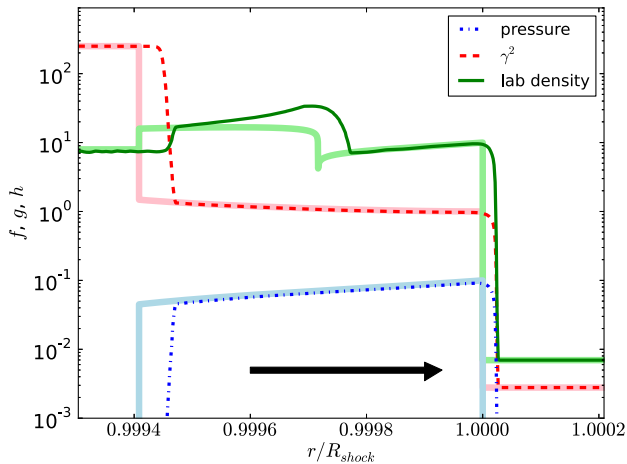}
\caption{Same as Fig. \ref{wind_before_figure}, now for an ISM environment.}
 \label{ISM_before_figure}
\end{figure}

In Figs. \ref{wind_before_figure} and \ref{ISM_before_figure} fluid profiles have been plotted for the case of sustained energy injection at time $t = T_{last}$ for $k = 0$ and $k = 2$. The radial shifts between the simulation and analytical fluid profiles that are clearly visible on the plots, are on the order of $10^{-2}$ percent of the FS radius, and translate to a few percent in terms of the total width of the RS-FS system. Velocity and pressure are reproduced excellently for both wind and ISM scenario's. In the wind case, the infinite spike in mass (where the hot fluid assumption breaks down as well, see section \ref{self-similar_blast_wave_subsection}) can not be reproduced numerically by definition. The reverse shock position is well captured by the analytical solution (accounting for the overall shift). The mass in the RS region (and not in the spike) exceeds the analytical value between 10 - 15 percent. The difference between densities for the simulation with 21 levels of refinement shown on the plot and one with 20 levels is far smaller, so this deviation is likely genuine for these explosion and medium parameters, although part of the explanation lies in the density change across the shock jump for the RS, which tends to be diffused numerically when not manually kept at peak refinement.

In the ISM case, the RS position lies a little ahead of its analytically prediction position (the difference being about twice the overall shift). The RS is captured more sharply, so the jump values for density match better than in the wind case. However, the RS region is smaller while containing the same amount of mass, leading again to higher densities in the simulation profile than analytically predicted. The singularity at the CD is inevitably diffused by the simulation. Nevertheless, both in the wind and ISM case, the profiles demonstrate how the approximate self-similar solution provides a reasonable prediction for the fluid behavior.

\begin{figure}
 \centering
  \includegraphics[width=\columnwidth]{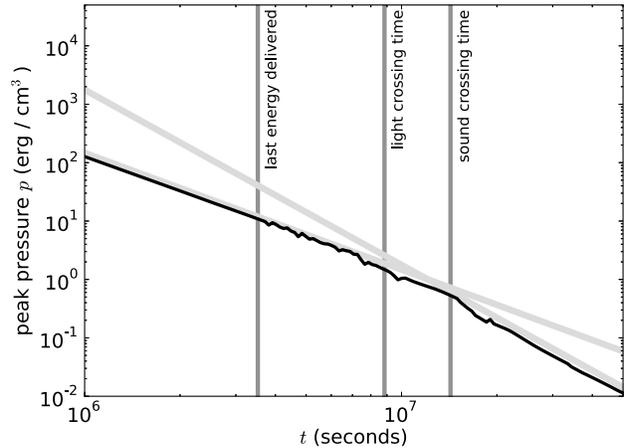}
\caption{Peak pressure $p$ for a typical blast wave in a stellar wind environment, with limited injection of energy for $T_{in} = 10^4$ s. The peak pressure is analytically expected to occur right behind the FS, and the solid black curve therefore shows the evolution of the FS. The thick grey lines denote the analytical sustained and impulsive energy injection BM solutions. The vertical lines indicate analytically calculated times of potential interest, from left to right: the point where the last energy is delivered across the RS, the point where this event would be communicated to the FS with the speed of light, the point where a sound wave communicating this event reaches the FS.}
 \label{wind_transition_figure}
\end{figure}

\begin{figure}
 \centering
  \includegraphics[width=\columnwidth]{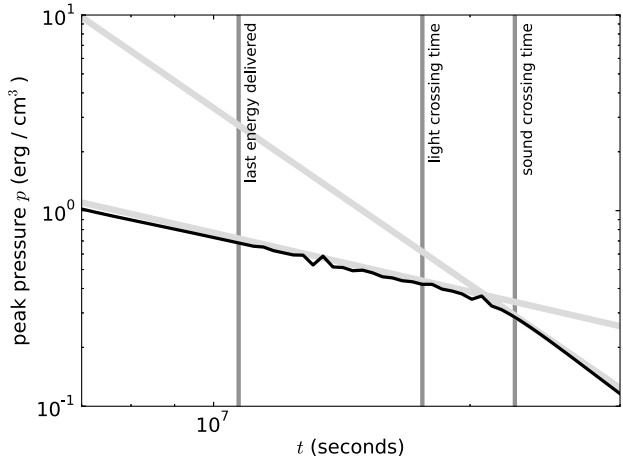}
\caption{Same as Fig. \ref{wind_transition_figure}, now for the ISM case.}
 \label{ISM_transition_figure}
\end{figure}

In Figs. \ref{wind_transition_figure} and \ref{ISM_transition_figure}, we turn to the time-evolution of the blast wave with limited energy injection. The peak pressure is analytically expected at the FS front in both cases, so plotting this quantity provides us with information on the behavior of the shock front. As expected by causality, both plots confirm that the shock front remains unaware of the cessation of energy injection through the RS until after a light crossing time across the RS-FS region. In both cases, the FS does start to deviate from sustained energy injection dynamics slightly before the theoretically predicted times. The discrepancy is most clearly seen in the ISM case, reflecting the fact that the simulation RS-CD-FS profiles are slightly thinner than analytically predicted, allowing a sound wave to get to the forward shock earlier.

\begin{figure}
 \centering
  \includegraphics[width=\columnwidth]{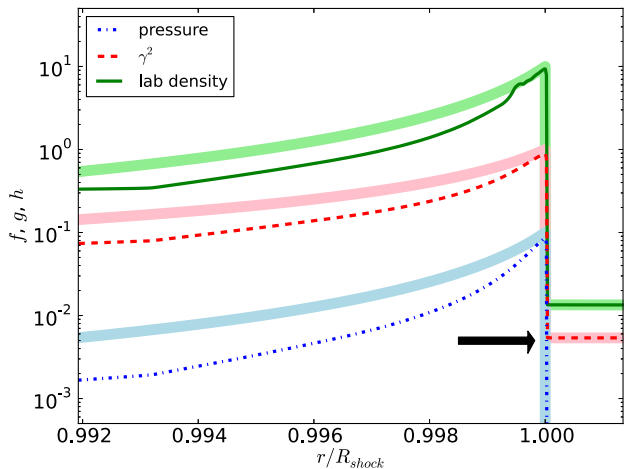}
\caption{Wind case profile, same as Fig. \ref{wind_before_figure}, now for finite energy injection with $T_{in} = 10^4$ s. and taken at the time where a sound wave leaving the reverse shock at the moment of last energy delivery reaches the FS, here $1.43 \times 10^{7}$ s.}
 \label{wind_after_figure}
\end{figure}

\begin{figure}
 \centering
  \includegraphics[width=\columnwidth]{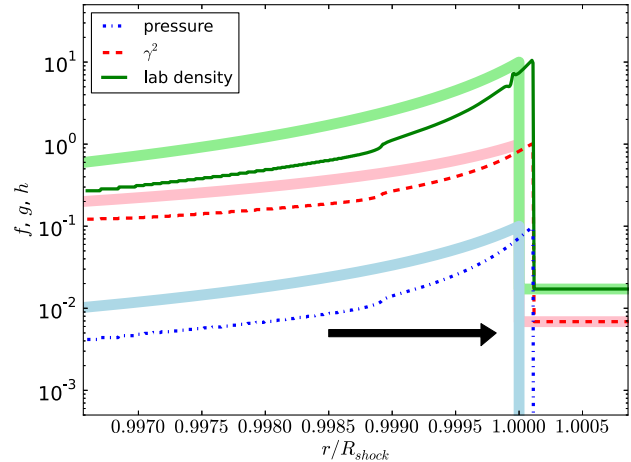}
\caption{ISM case profile, same as Fig. \ref{ISM_before_figure}, now for finite energy injection with $T_{in} = 10^4$ s. and taken at the time where a sound wave leaving the reverse shock at the moment of last energy delivery reaches the FS, here $2.26 \times 10^{7}$ s.}
 \label{ISM_after_figure}
\end{figure}

The implications of this for the post-transition fluid profile are shown in Figs. \ref{wind_after_figure} and \ref{ISM_after_figure}. A small overall shift remains in the ISM case, but no longer in the wind case. The remaining large-scale feature is the FS and the fluid profiles for $p$, density and $\gamma$ are all moving to their new asymptotic self-similar values. When looking at the density profile (in green), it can be seen how far this transition is along and the steep drop away from the analytical solution, slightly behind the shock front and seen for both wind and ISM, marks a newly formed CD, separating external fluid shocked since the cessation of energy injection was communicated to the front, from previously shocked external medium. Its existence separately confirms that the cessation of energy injection is communicated slightly ahead of the analytically predicted time, given that the snapshot times were chosen to match this predicted time. Across this new CD, pressure and velocity remain continuous, as they should. The ISM case also shows a newly formed RS still within the plot, which runs into the old FS region and communicates backwards the existence of the forward shock (now in a new decelerating phase consistent with impulsive energy injection).

In all, one can estimate the point where the transition is completed fully, to the extent that even the fluid profiles match the impulsive BM solution, as follows. First, take the sound crossing time, then allow for the newly formed CD to advect with the flow until the approximate `back' of the blast wave at $\chi_{B}$. The latter takes a factor $X = \chi_B^{1/(4-k)}$ to complete (cf. eq. \ref{impulsive_crossing_time_equation}). Following constant energy injection with $q = 0$, this implies that, according to our estimate, the transition is completed at $T_{comp} = 2.11 \times 1.15$ $T_{last}$ for the ISM case and at $T_{comp} = 4.05 \times 1.22$ $T_{last}$ for the wind case, where $T_{last}$ the time when the last injected energy crosses the RS, given by eq. \ref{t_and_t_in_equation}.

\section{Light curve predictions for energy injection flows}
\label{light_curve_section}

\begin{figure}
 \centering
  \includegraphics[width=\columnwidth]{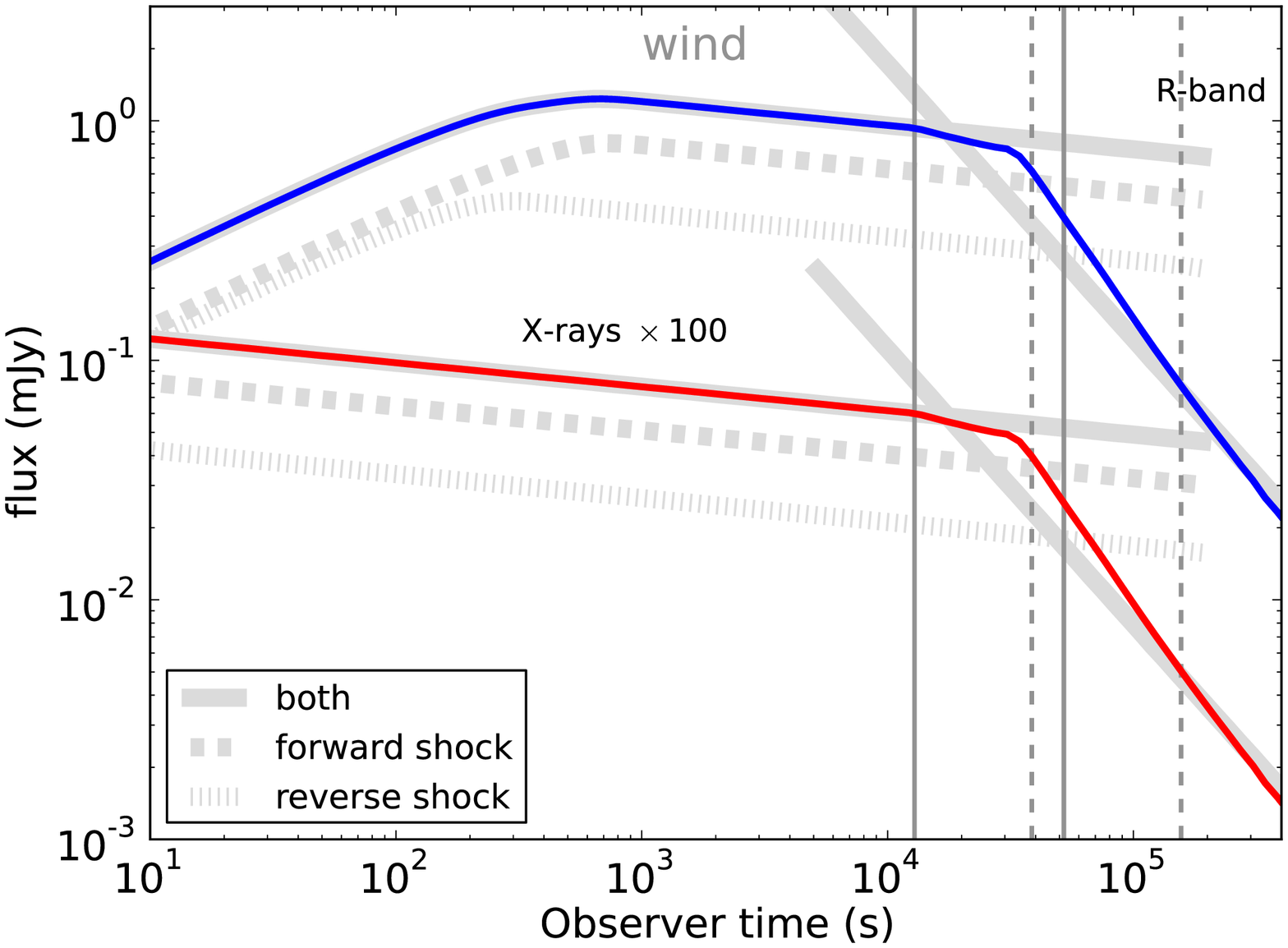}
  \includegraphics[width=\columnwidth]{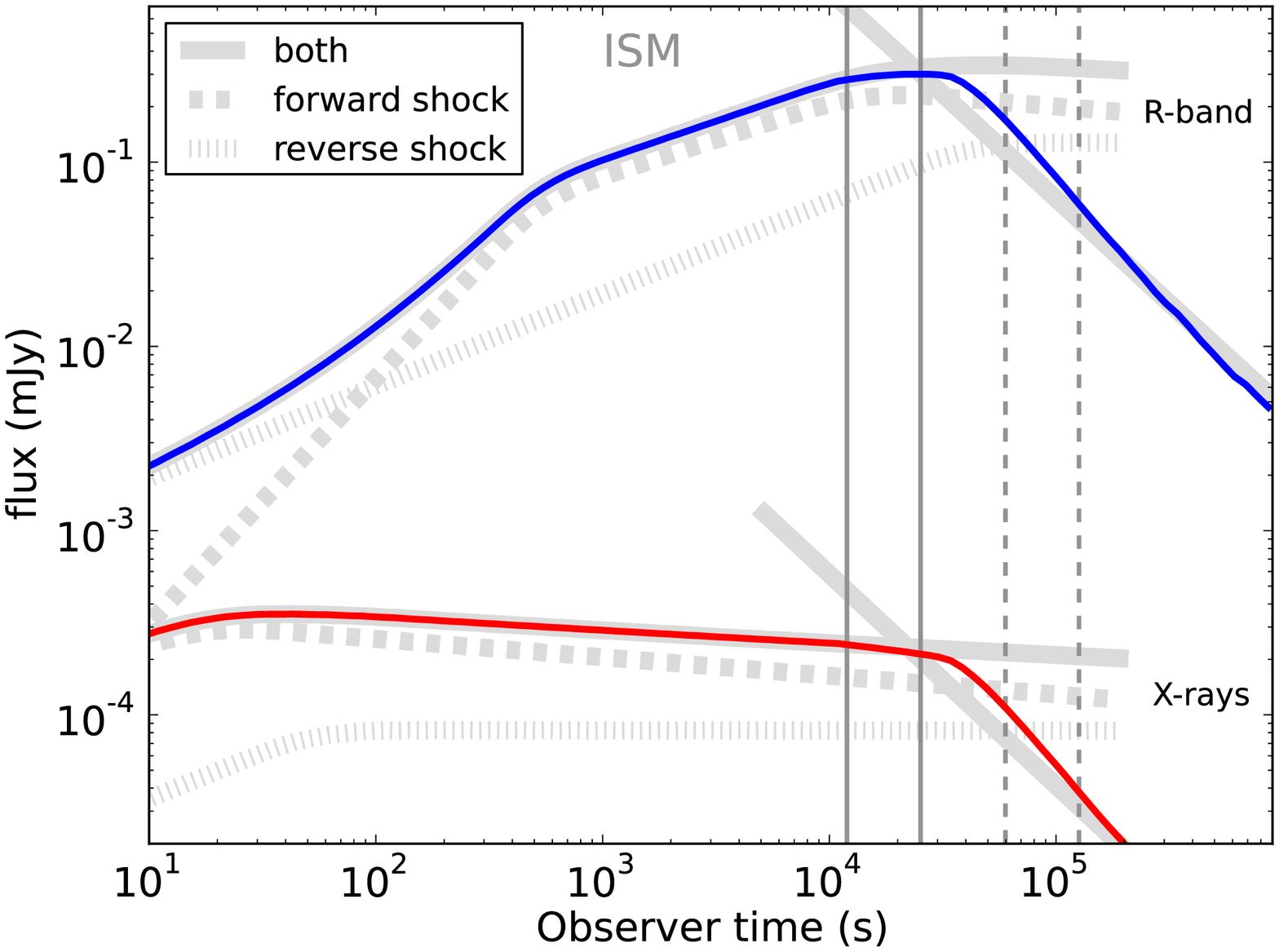}
\caption{Optical ($\nu = 4.56 \times 10^{14}$ Hz `R-band', upper blue curve) and X-ray ($\nu = 3.63 \times 10^{17}$ Hz, lower red curve) light curves for typical wind (top plot) and ISM (bottom plot) scenario's. Thick light grey curves plot the analytical solutions for both sustained and impulsive energy injection. Thick dashed light grey curves plot the forward shock region emission only, thick dotted light grey curves the reverse shock region only. From left to right, the grey vertical lines indicate (1) the arrival time of emission from the jet back, sent at the moment when the last injected energy passes through the RS and (2) the arrival time of emission from the jet front, sent at the moment where the last injected energy arrival is communicated to the front via a sound wave. The solid vertical lines are arrival times of emission along the jet axis for these two events, the dashed vertical lines correspond to arrival times of emission from an angle $\theta = 1 / \gamma$.}
 \label{light_curves_typical_figure}
\end{figure}

\begin{table*}
\centering
\begin{tabular}{|llll|}
\hline
 $F$ or $\nu$ & scalings & $\kappa$ (energy) & $\lambda$ (density) \\
\hline
 & & & \\
\normalsize $ \displaystyle F_{peak,I}$ & \normalsize $\displaystyle \frac{(1+z)^{\frac{k-8}{2(k-4)}}}{d_L^2} \xi_N \epsilon_B^{\frac{1}{2}} \left( n_{ref} R_{ref}^k \right)^{\frac{-2}{k-4}} E_{iso}^{\frac{-8+3k}{2(k-4)}} t_{obs}^{\frac{k}{2(k-4)}}$ & $\displaystyle  \kappa^{\frac{3(k-2)}{2(k-3)}}$ & \normalsize $\displaystyle \lambda^{\frac{-3}{2(k-3)}}$ \\ [15pt]

 \normalsize $\displaystyle \nu_{m,I}$ & \normalsize $\displaystyle (1+z)^{\frac{1}{2}} \xi_N^{-2} \epsilon_e^2 \epsilon_B^{\frac{1}{2}} E_{iso}^{\frac{1}{2}} t_{obs}^{-\frac{3}{2}} $ & \normalsize $\displaystyle \kappa^{\frac{k}{2(k-3)}}$ & \normalsize $\displaystyle \lambda^{\frac{-3}{2(k-3)}} $ \\ [15pt]

 \normalsize $\displaystyle \nu_{c,I}$ & \normalsize $\displaystyle (1+z)^{\frac{4+k}{2(k-4)}} \epsilon_B^{-\frac{3}{2}} \left( n_{ref} R_{ref}^k \right)^{\frac{4}{k-4}} E_{iso}^{\frac{4-3k}{2(k-4)}} t_{obs}^{\frac{4-3k}{2(k-4)}} $ & \normalsize $\displaystyle \kappa^{\frac{4-3k}{2(k-3)}}$ & \normalsize $\displaystyle \lambda^{\frac{5}{2(k-3)}}$ \\  [15pt]

\hline
 & & & \\
\normalsize $\displaystyle F_{peak,FS}$ & \normalsize $\displaystyle \frac{\left( 1+z \right)^{\frac{k-8}{2(k-4)}}}{d_L^2} \xi_N \epsilon_B^{\frac{1}{2}} \left(n_{ref} R_{ref}^k \right)^{\frac{-2}{k-4}} \left( E_{iso} T_{in, \oplus}^{-(1+q)} \right)^{\frac{-(8-3k)}{2(k-4)}} t_{obs}^{\frac{4k - 8 -8q + 3kq}{2(k-4)}}$  &  \normalsize same as $I$ &  \normalsize same as $I$ \\  [15pt]

 \normalsize $\displaystyle \nu_{m,FS}$ & \normalsize $\displaystyle (1+z)^{\frac{1}{2}} \xi_N^{-2} \epsilon_e^2 \epsilon_B^{\frac{1}{2}} \left( E_{iso} T_{in, \oplus}^{-(1+q)} \right)^{\frac{1}{2}} t_{obs}^{\frac{q-2}{2}}$ & \normalsize same as $I$ & \normalsize same as $I$ \\  [15pt]

 \normalsize $\displaystyle \nu_{c,FS}$ & \normalsize $\displaystyle (1+z)^{\frac{k+4}{2(k-4)}} \epsilon_B^{-\frac{3}{2}} \left( n_{ref} R_{ref}^k \right)^{\frac{4}{k-4}} \left( E_{iso} T_{in, \oplus}^{-(1+q)} \right)^{\frac{4-3k}{2(k-4)}} t_{obs}^{\frac{-(2+q)(3k-4)}{2(k-4)}}$ & \normalsize same as $I$ & \normalsize same as $I$ \\  [15pt]

\hline
 & & & \\
 \normalsize $F_{peak,RS}$ &  \normalsize $\displaystyle \frac{\left( 1 + z \right)^{\frac{3k-14}{2(k-4)}}}{d_L^2} \eta^{-1} \xi_N \epsilon_B^{\frac{1}{2}} \left( n_{ref} R_{ref}^k \right)^{\frac{-1}{k-4}} \left( E_{iso} T_{in, \oplus}^{-(1+q)} \right)^{\frac{3k-10}{2(k-4)}} t_{obs}^{\frac{2k-4-10q+3kq}{2(k-4)}}$ &  \normalsize same as $I$ &  \normalsize same as $I$ \\  [15pt]

 \normalsize $\nu_{m,RS}$ &  \normalsize $\displaystyle (1+z)^{\frac{8-3k}{2(k-4)}} \eta^2 \xi_N^{-2} \epsilon_e^2 \epsilon_B^{\frac{1}{2}} \left( n_{ref} R_{ref}^k \right)^{\frac{-2}{k-4}} \left( E_{iso} T_{in, \oplus}^{-(1+q)} \right)^{\frac{k}{2(k-4)}} t_{obs}^{\frac{k(q+2)}{2(k-4)}}$ &  \normalsize same as $I$ &  \normalsize same as $I$ \\  [15pt]

 \normalsize $\nu_{c,RS}$ & \normalsize same as $FS$ & \normalsize same as $FS$, $I$ & \normalsize same as $FS$, $I$ \\ [15pt]

\hline
\end{tabular}
\caption{Flux scalings for the characteristic quantities of the synchrotron spectrum. $\kappa$ and $\lambda$ are defined as in \protect\cite{vanEerten2012boxfit, vanEerten2012scalings}.}
\label{characteristics_scalings_table}
\end{table*}

\begin{table*}
\centering
\begin{tabular}{|ll|}
\hline
regime & scalings \\
\hline
 & \\
\normalsize $F_{D, FS}$ & \normalsize $d_L^{-2} (1+z)^{\frac{k-10}{3(k-4)}} \xi_N^{\frac{5}{3}} \epsilon_e^{-\frac{2}{3}} \epsilon_B^{\frac{1}{3}} \left( n_{ref} R_{ref}^k \right)^{\frac{-2}{k-4}} \left( E_{iso} T_{in, \oplus}^{-(q+1)} \right)^{\frac{4k-10}{3(k-4)}} t_{obs}^{\frac{7k+4kq-10q-16}{3(k-4)}} \nu^{\frac{1}{3}}$ \\ [15pt]

\normalsize $F_{E, FS}$ & \normalsize $d_L^{-2} (1+z)^{\frac{k-14}{3(k-4)}} \xi_N \epsilon_B \left( n_{ref} R_{ref}^k \right)^{\frac{-10}{3(k-4)}} \left(E_{iso} T_{in, \oplus}^{-(q+1)} \right)^{\frac{6k-14}{3(k-4)}} t_{obs}^{\frac{9k+6kq-14q-16}{3(k-4)}} \nu^{\frac{1}{3}}$\\ [15pt]

\normalsize $F_{F, FS}$ & \normalsize $d_L^{-2} (1+z)^{\frac{3}{4}} \xi_N \epsilon_B^{-\frac{1}{4}} \left( E_{iso} T_{in, \oplus}^{-(q+1)} \right)^{\frac{3}{4}} t_{obs}^{\frac{2+3q}{4}} \nu^{\frac{-1}{2}}$ \\ [15pt]

\normalsize $F_{G, FS}$ & \normalsize $d_L^{-2} (1+z)^{\frac{k-12-4p+pk}{4(k-4)}} \xi_N^{2-p} \epsilon_e^{p-1} \epsilon_B^{\frac{1+p}{4}} \left( n_{ref} R_{ref}^k \right)^{\frac{-2}{k-4}} \left( E_{iso} T_{in, \oplus}^{-(1+q)} \right)^{\frac{5k-12-4p+pk}{4(k-4)}} t_{obs}^{\frac{10k + 5kq - 12q - 24 - 4pq - 2pk + 8p + pkq}{4(k-4)}} \nu^{\frac{1-p}{2}}$ \\ [15pt]

\normalsize $F_{H, FS}$ & \normalsize $d_L^{-2} (1+z)^{\frac{2+p}{4}} \xi_N^{2-p} \epsilon_e^{p-1} \epsilon_B^{\frac{p-2}{4}} \left( E_{iso} T_{in, \oplus}^{-(1+q)} \right)^{\frac{2+p}{4}} t_{obs}^{\frac{4+2q-2p+pq}{4}} \nu^{-\frac{p}{2}}$ \\ [15pt]
\hline

 & \\
\normalsize $F_{D, RS}$ & \normalsize $d_L^{-2} (1+z)^{\frac{3k-13}{3(k-4)}} \eta^{-\frac{5}{3}} \xi_N^{\frac{5}{3}} \epsilon_e^{-\frac{2}{3}} \epsilon_B^{\frac{1}{3}} \left( n_{ref} R_{ref}^k \right)^{\frac{-1}{3(k-4)}} \left(E_{iso} T_{in, \oplus}^{-(1+q)} \right)^{\frac{4k-15}{3(k-4)}} t_{obs}^{\frac{2k+4kq-15q-6}{3(k-4)}} \nu^{\frac{1}{3}}$ \\ [15pt]

\normalsize $F_{E, RS}$ & \normalsize $d_L^{-2} (1+z)^{\frac{k-11}{3(k-4)}} \eta^{-1} \xi_N \epsilon_B \left( n_{ref} R_{ref}^k \right)^{\frac{-7}{3(k-4)}} \left( E_{iso} T_{in, \oplus}^{-(q+1)} \right)^{\frac{-17+6k}{3(k-4)}} t_{obs}^{\frac{-17q+6k-10+6kq}{3(k-4)}} \nu^{\frac{1}{3}}$ \\ [15pt]

\normalsize $F_{F, RS}$ & \normalsize $d_L^{-2} (1+z)^{\frac{3k-8}{4(k-4)}} \eta^{-1} \xi_N \epsilon_B^{-\frac{1}{4}} \left( n_{ref} R_{ref} \right)^{\frac{1}{k-4}} \left( E_{iso} T_{in, \oplus}^{-(1+q)} \right)^{\frac{-16+3k}{4(k-4)}} t_{obs}^{\frac{-16q-2k+3kq}{4(k-4)}} \nu^{-\frac{1}{2}}$ \\ [15pt]

\normalsize $F_{G, RS}$ & \normalsize $d_L^{-2} (1+z)^{\frac{-20+5k+8p-3pk}{4(k-4)}} \eta^{p-2} \xi_N^{2-p} \epsilon_e^{p-1} \epsilon_B^{\frac{1+p}{4}} \left( n_{ref} R_{ref}^k \right)^{\frac{-p}{k-4}} \left(E_{iso} T_{in, \oplus}^{-(1+q)} \right)^{\frac{-20+5k+pk}{4(k-4)}} t_{obs}^{\frac{-20q+2k -8+5kq+2kp+kqp}{4(k-4)}} \nu^{\frac{1-p}{2}}$ \\ [15pt]

\normalsize $F_{H, RS}$ & \normalsize $d_L^{-2} (1+z)^{\frac{-32+10k+8p-3pk}{4(k-4)}} \eta^{p-2} \xi_N^{2-p} \epsilon_e^{p-1} \epsilon_B^{\frac{p-2}{4}} \left(n_{ref} R_{ref}^k \right)^{\frac{2-p}{k-4}} \left( E_{iso} T_{in, \oplus}^{-(1+q)} \right)^{\frac{-16+2k+pk}{4(k-4)}} t_{obs}^{\frac{-16q-4k+2kq+2pk+pkq}{4(k-4)}} \nu^{-\frac{p}{2}}$ \\ [15pt]
\hline
\end{tabular}
\caption{Flux scalings for the fluxes in the various spectral regimes.}
\label{flux_scalings_table}
\end{table*}

In order to link the dynamical energy injection model to GRB afterglow observations, it can be combined with a synchrotron radiation module. In the afterglow phase, synchrotron emission, by shock-accelerated electrons interacting with local small-scale magnetic fields (presumably also shock-generated), is typically the dominant emission mechanism. In the standard approach (e.g. \citealt{Meszaros1997, Wijers1997, Meszaros1998, Sari1998, Wijers1999, Rhoads1999, Granot1999, Gruzinov1999}) a fraction $\epsilon_e \sim 0.1$ of the local energy density is assumed to reside in the accelerated electron population, a fraction $\epsilon_B \sim 0.01$ in the magnetic field. A fraction $\xi_N \sim 1$ of the available electrons are assumed accelerated into a power law distribution over energies with slope $-p \sim -2.2$ (see e.g. \citealt{Curran2009, Ryan2013, Ryan2014}).

In brief, the peak synchrotron flux $F_{peak}$ in the observer frame is proportional to the number of radiating particles and the (comoving) field strength $B$ according to $F_{peak} \propto \Gamma^2 \xi_N n B V$, where the volume of the thin shell $V \propto R^3 / \Gamma^2$. The synchrotron break frequency $\nu_m$ in the observer frame is given by $\nu_m \propto \Gamma \gamma_m^2 B$, with $\gamma_m \propto \epsilon_e e / (\xi_N n)$ (the ratio between comoving energy density and number density). In these equations $B^2 \propto \epsilon_B e$. The cooling break frequency in the observer frame $\nu_c \propto \Gamma \gamma_c^2 B$, with $\gamma_c \propto \gamma / (B^2 t)$. Including $\nu_m$ and $\nu_c$, but ignoring the synchrotron self-absorption characteristic frequency typically associated with radio emission, the different orderings of observer frequency $\nu$, $\nu_m$ and $\nu_c$ lead to the observation of different spectral regimes, which in this study are labeled according to:
\begin{eqnarray}
F_D \equiv F_{peak} (\nu / \nu_m)^{1/3} & : & \nu < \nu_m < \nu_c, \nonumber \\
F_E \equiv F_{peak} (\nu / \nu_c)^{1/3} & : & \nu < \nu_c < \nu_m, \nonumber \\
F_F \equiv F_{peak} (\nu / \nu_c)^{-1/2} & : & \nu_c < \nu < \nu_m, \nonumber \\
F_G \equiv F_{peak} (\nu / \nu_m)^{(1-p)/2} & : & \nu_m < \nu < \nu_c, \nonumber \\
F_H \equiv F_{peak} (\nu_c / \nu_m)^{(1-p)/2} (\nu / \nu_c)^{-p/2} & : & \nu_m, \textrm{ } \nu_c < \nu. \nonumber
\end{eqnarray}
This follows the same naming conventions as \cite{Granot2002, vanEerten2009}.

Both the self-similar solutions and the simulations are used as input for the linear radiative transfer approach to synchrotron emission described in \cite{vanEerten2009, vanEerten2010transrelativistic, vanEerten2010offaxis}. In \cite{vanEerten2010offaxis} the exact equations of the implementation used in this paper can be found, and the approach to electron cooling from that paper is applied as well, where we treat the fluid as a single steady-state plasma with a global cooling time (subtleties regarding electron cooling are discussed in \citealt{vanEerten2013proceedings} and section \ref{cooling_subsection} of this paper).

The dependencies of the flux equations on the model parameters can be calculated analytically for each spectral regime and are tabulated in tables \ref{characteristics_scalings_table} and \ref{flux_scalings_table}. In these tables, fluxes, frequencies and times are all expressed in the observer frame. In order to translate the observer time $t_{obs}$, the energy injection duration $T_{in, \oplus}$ and the peak flux to the burster frame where redshift $z = 0$, they need to be divided by $(1+z)$. Frequencies need to be multiplied by $(1+z)$. Note that in \cite{vanEerten2012scalings}, times and frequencies are expressed in the burster frame in order to simplify the equations; the equations here are directly applicable to observations.

Not included in the tables \ref{characteristics_scalings_table} and \ref{flux_scalings_table} are the numerical prefactors that fix the absolute flux levels. These have been deferred to appendix \ref{heuristic_flux_appendix} and can be included if one wishes to directly compare model predictions to data.

As described for impulsive energy injection in \cite{vanEerten2012scalings}, the dynamical scale invariance between total explosion energies and circumburst densities \citep{vanEerten2012boxfit} carries over to light curves, albeit differently for each spectral regime. The $\kappa$ and $\lambda$ columns of table \ref{characteristics_scalings_table}, showing the energy and density scalings for the impulsive energy injection stage (drawn from \citealt{vanEerten2013boostedcurves}), remain unchanged when considering sustained energy injection instead. A simulation-based evolution curve for any given characteristic quantity, say $\nu_m$, can be scaled to different energies and densities even when it includes a transition from energy injection to impulsive injection: plateau flux and transition time just scale along. Scaling up the total explosion energy without co-scaling $T_{in}$, requires adding an extra dimension to parameter space, as does including different $q$ values.

\subsection{Application to typical afterglow parameters}

The plots of Fig. \ref{light_curves_typical_figure} show optical and X-ray light curves for the typical values of the model parameters discussed in section \ref{typical_values_section}. In addition I have taken a redshift $z = 2.23$ (the average \emph{Swift} sample redshift in 2009, see \citealt{Evans2009}) and luminosity distance $d_L = 5.6 \times 10^{28}$ cm, but took the value $10^4$ for $T_{in}$ s. as referring to the burster frame duration, such that $T_{in} (1+z) = 3.23 \times 10^4$ s. The plots show light curves generated both directly from the analytical solutions for the dynamics and from the numerical simulations that cover the transition stage. For the early emission from the simulations, from before they numerically established the expected self-similar injection profile, analytical fluid profiles were used.

The light curves of Fig. \ref{light_curves_typical_figure} demonstrate a few key points:

\begin{enumerate}
  \item \textbf{The reverse shock contribution can be significant or dominant.} In our `typical' scenario's we have assumed the same magnetization for both regions. Even so, the RS flux dominates the FS flux in the optical for $10^2$ s in the ISM case, and both flux levels are comparable in the wind case. The magnetization of the FS region is a result from magnetic field generation at the shock front, and to a (presumably) lesser extent, compression of the ambient magnetic field. The original ejecta (i.e. the RS region) can be magnetized to a far higher degree (see e.g. \citealt{ZhangBing2005, Giannios2008, Mimica2009}). As a result of this difference in magnetization, emission from the RS region can easily be made to dominate the overall flux output, especially when the FS region magnetization is, in turn, weak (see e.g. \citealt{Kumar2010, Santana2013}) \\

  \item \textbf{When the flux contributions from FS and RS region are comparable, the light curve slope will reflect both contributions.} Examples of this are given by the X-ray and optical emission for the typical ISM case and the early time optical emission for the typical wind case.\\

  \item \textbf{The transition between regimes in the light curve occurs around when cessation of energy injection is communicated to the shock front.} A number of grey vertical lines in Fig. \ref{light_curves_typical_figure} indicate characteristic times for the typical scenario's. A deviation from the sustained energy injection asymptote is first seen when the last of the energy is delivered across the RS. Due to differences in arrival times between different emission angles, the initial change is small. The transition nears completion when the high angle emission is seen that is emitted at time the sound wave from the RS reaches the FS. Here the upper angle is defined by the width of the beaming cone (i.e. $\theta \sim 1 / \gamma$; for narrowly collimated ejecta one should use $\theta_0$ instead). The equations for the two arrival times using the on-axis emission are:
  \begin{eqnarray}
  t_{0, \oplus} & = & (1 + z) T_{in} / \chi_{RS}, \nonumber \\
  t_{1, \oplus} & = & (1 + z) T_{in} X / \chi_{RS},
  \end{eqnarray}
where $X$ the corresponding factor from table \ref{soundwave_table}. For emission from the edge of the beaming cone, we have:
  \begin{eqnarray}
  t_{0, \oplus} & = & (1 + z) T_{in} (2m + 3)/ \chi_{RS}, \nonumber \\
  t_{1, \oplus} & = & (1 + z) T_{in} X (2m + 3) / \chi_{RS}.
  \end{eqnarray}
  These differences in arrival times between on- and off-axis emission are related to the well-known \emph{curvature} effect, putting a limit on the steepness of light curve decay even if the emission where suddenly switched off at the source (see e.g. \citealt{KumarPanaitescu2000}). Once the highest angle emission from the jet edges has arrived, the subsequent drop in flux can be arbitrarily steep. \\ 

  \item \textbf{The optical light curve peak does not necessarily mark the onset of the deceleration stage of massive ejecta.} This point was also raised by \cite{Leventis2014}. In Fig. \ref{light_curves_typical_figure} this is illustrated by the peak at $\sim 500$ s, which is due to a spectral transition (the passing of $\nu_m$ through the observer band), rather than the onset of deceleration.\\
\end{enumerate}  

\begin{table}
\centering
\resizebox{\columnwidth}{!}{
\begin{tabular}{|llll|}
regime & $p$, $k$ & $k = 0$, $\Delta p$ & $k = 2$, $\Delta p$ \\
\hline
 $F_{D, FS}$ & $\frac{k-2}{k-4}$ & $0.5$ & $0$ \\

$F_{E, FS}$ & $\frac{3k - 2}{3(k-4)}$ &  $0.167$ & $-0.67$ \\

$F_{F,FS}$ & $-\frac{1}{4}$ & $-0.25$ & $-0.25$ \\

$F_{G,FS}$ & $\frac{5k - 12 + 12p - 3pk}{4(k-4)}$ & $-(0.9 + 0.75 \Delta p)^b$ & $-(1.4 + 0.75 \Delta p)$ \\ 

$F_{H,FS}$ & $\frac{2 - 3p}{4}$ &  $-(1.15 + 0.75 \Delta p)^b$ &  $-(1.15 + 0.75 \Delta p)^b$ \\

\hline

 $F_{D,RS}$ & $\frac{-2k+9}{3(k-4)}$ & $-0.75^a$ &  $-0.833^a$ \\

$F_{E,RS}$ & $\frac{7}{3(k-4)}$ & $-0.583$ & $-1.167$ \\

 $F_{F,RS}$ & $\frac{-5k+16}{4(k-4)}$ & $-1^b$ &  $-0.75^a$ \\

 $F_{G,RS}$ & $\frac{12-3k+pk}{4(k-4)}$ &  $-0.75^a$ & $-(1.3 + 0.25 \Delta p)$ \\

 $F_{H,RS}$ & $\frac{16-6k+pk}{4(k-4)}$ & $-1^b$ &  $-(1.05 + 0.25 \Delta p)^b$ \\
\hline
\end{tabular}}
\caption{Relation $F \propto t^{\ldots}$ between flux $F$ and time $t$, at the point where $t = T_{in}$, for the different spectral regimes. In addition to the general case with unspecified $p$ and $k$, the ISM and wind cases are listed separately. In the last two columns, $\Delta p \equiv p - 22 / 10$ is used to emphasize the value around $p \approx 2.2$. Superscript \emph{a} marks those entries consistent with the $F_b - t_b$ correlation for optical emission, superscript \emph{b} marks those consistent with the X-ray $F_b - t_b$ correlation. For both we assumed the range of $p$ to be $2.07 - 2.51$ \citep{Ryan2014}.}
\label{F_T_table}
\end{table}

In addition to the points made above, the flux equations also demonstrate the following:

\begin{enumerate}
\setcounter{enumi}{4}

 \item \textbf{RS emission in a wind environment is usually in the fast cooling regime, RS emission in a homogeneous environment in the slow cooling regime}. For slow cooling $\nu_m < \nu_c$, for fast cooling $\nu_c < \nu_m$. In the typical ISM case, $\nu_{m, RS}$ will remain fixed at $1.2 \times 10^{12}$ Hz, while $\nu_{c,RS}$ will decrease according to $\nu_{c,RS} \propto t_{obs}^{-1}$ and meet $\nu_m$ at $1.3 \times 10^7$ s $\gg T_{in}$, passing through $1.5$ keV at 43 s and the V band at $2.8 \times 10^4$ s. This implies that X-ray afterglow light curves (e.g. from \emph{Swift}'s XRT) will have $\nu > \nu_c, \nu_m$, while optical light curves (e.g. from \emph{Swift}'s UVOT) have $\nu_m < \nu < \nu_c$, consistent with the spectral slopes typically found in both regimes (see e.g. \citealt{Liang2008, Racusin2009, Li2012}). It also implies an observationally motivated upper limit on $\eta$, given that $\nu_{m,RS} \propto \eta^2$ will easily lead to $\nu_m > \nu$ in the optical, which seems hard to reconcile with reported spectral slopes. For otherwise standard parameters, this limit lies around $\eta \sim 6 \times 10^3$, although this can be offset by a strong decrease in circumburst density or magnetic field strength, according to $\nu_{m,RS} \propto \epsilon_B^{1/2} n_{ref}^{1/2}$, both parameters that are poorly constrained by observations.
 
 In a wind environment, the situation is different. Now, with $q=0$, it is $\nu_{m,RS}$ that decays according $\nu_{m,RS} \propto t_{obs}^{-1}$, while $\nu_{c,RS} \propto t_{obs}^{1}$. Typically, $\nu_{m,RS}$ crosses the optical bands around 90 s (which can be mistaken for the onset of deceleration, as mentioned previously). Both frequencies meet at $3 \times 10^3$ s, close to the end of the plateau. In this case the strong dependency of $\nu_{m,RS} \propto \eta^2$ means that fast cooling will persist longer, and well past the plateau phase, for larger values of $\eta$. If for the X-rays we insist on $\nu > \nu_c, \nu_m$ at least from $10^2$ s onward, this implies $\eta < 8 \times 10^3$ for otherwise typical wind parameters.
 
 For $p \sim 2.2$, the shape of the spectrum looks very similar regardless of which critical frequency is highest. Add to this that the RS emission will dissipate post-plateau, leading to a light curve eventually dictated by an impulsive injection FS, and it follows that conclusions about the nature of the spectrum (i.e. slow vs. fast cooling) at late times can not be automatically extrapolated to early plateau times, such that it becomes hard to dismiss out of hand a fast cooling scenario during the plateau phase.
 
 Finally, combining the observed temporal slopes for X-rays and optical with the flux equations presented here, one finds that the conclusion remains unaltered that observationally $q < 0$ (like -0.5 for a FS analysis, \citealt{ZhangBing2006}). In order for the light curve to decay for a wind scenario in fast-cooling regime $F$, $q < -0.4$. \\

  \item \textbf{The observational $F_b-t_b$ correlations are fully supported by long-term energy injection.} This issue has been discussed by \cite{Leventis2014} for homogeneous environments. From measurements of the optical flux $F_b$ at the observed end time $t_b$ of the plateau phase, a correlation $F_b \propto t_b^{-0.78 \pm 0.08}$, has been found \citep{PanaitescuVestrand2011, Li2012}. In X-rays, a similar correlation but with different negative index $1.07^{+0.09}_{-0.14}$ has been reported \citep{Dainotti2008, Dainotti2010, Dainotti2013, Grupe2013, Margutti2013}. As the flux equations of table \ref{flux_scalings_table} demonstrate, these kind of correlations naturally emerge in a $q$-independent fashion when the flux is measured at time $t = t_b \sim T_{in}$. At this point in time, all energy is added to the blast wave and the relevant parameters are no longer $L_0$, $q$ and $T_{in}$, but $E_{iso}$ instead. In table \ref{F_T_table}, the dependencies of flux on time when $t = T_{in}$ are listed for all spectral regimes. For the FS emission, these are identical to the impulsive energy injection flux time dependencies; something which naturally follows from dimensional analysis using $E_{iso}$ and $t$. For the RS emission, the time dependencies are different due to additional dimensionless factors introduced by the density ratio between FS and RS region (cf. eq. \ref{density_ratio_equation}). For the correlation to follow from this model, no implicit cross-correlations between model parameters can can exist. $E_{iso}$ and $T_{in}$ need to be independent parameters (i.e. instead of $L_0$ and $T_{in}$). Uncertainties and the intrinsic range of all model parameters ($E_{iso}$, $\epsilon_B$, $n_0$, etc.) will lead to scatter in the correlation, but not impact its slope.
  
For X-rays we can combine the observed correlation with the observation by \cite{Racusin2009} that during the plateau phase the spectral slope is broadly clustered around $-1$ (see the distribution for `segment II' in Fig. 2 of the cited paper). This implies that $F (t \sim T_{in}) \propto t^{-1.07} \nu^{-1.}$, most consistent with $\nu > \nu_{m}, \nu_c$. It should be noted here that the spectral transition across the cooling break is very smooth \citep{Granot2002, vanEerten2009, Uhm2014}, so the asymptotic power law limit for the slope might not actually be applicable. Regardless, a spectral slope of -1 seems hard to reconcile with the expected slope $\sim -0.5$ expected when $\nu < \nu_c$. This points us towards the entries for $F_{H}$ in table \ref{F_T_table}, where we see  that both FS and RS can account for the observed X-ray correlation, both in the wind and ISM case. Out of the four possibilities, RS shock emission for a blast wave running into a stellar wind environment gives a prediction marginally closer to the observed value than the others. It still depends on $p$, but this dependency is only weak, $\Delta p / 4$, where $\Delta p \equiv p - 2.2$

None of the FS emission regions supports the optical $F_b - t_b$ correlation within 1 $\sigma$, although $F_{G}$ in the ISM case gets close within $2 \sigma$. According to table \ref{F_T_table}, the RS options for the ISM case are $\nu < \nu_m < \nu_c$ ($D$) and $\nu_m < \nu < \nu_c$ ($G$). For the wind case, they are $\nu < \nu_m < \nu_c$ ($D$) and $\nu_c < \nu < \nu_m$ ($F$). The measured spectral slope in the optical is usually negative (see e.g. the collected results in \citealt{Liang2008, Li2012} and references therein; although these values might not fully reflect early plateau time values and evolution of the spectral slope during early times), which, if taken at face value, would rule out $D$ and leave us with either the slow cooling (ISM) or fast cooling option (wind). Both would yield a negative spectral slope of $~0.5$ (assuming $p \sim 2.2$). The reported values from \cite{Li2012} are often consistent with this slope, but also often lie higher, around $0.75$. The latter implies either a larger value for $p$, or that $\nu$ approaches the (smooth) spectral break towards spectral regime $H$ (i.e. approaches $\nu_c$ for ISM or $\nu_m$ for wind).

At this point, the optical correlation therefore seems weakly suggestive of the RS being the dominant emission region. However, it remains to be tested whether this holds once the flux equations from the model are tested in a more realistic fashion, where biases and error margins due to instrumental systematics or underlying population distributions realized in nature are fully taken into account. This is the topic of a follow-up study \citep{vanEerten2014correlations}.

\end{enumerate}

\subsection{Application to GRB 090515 and 120521A}
\label{data_application_subsection}

\begin{figure}
 \centering
  \includegraphics[width=\columnwidth]{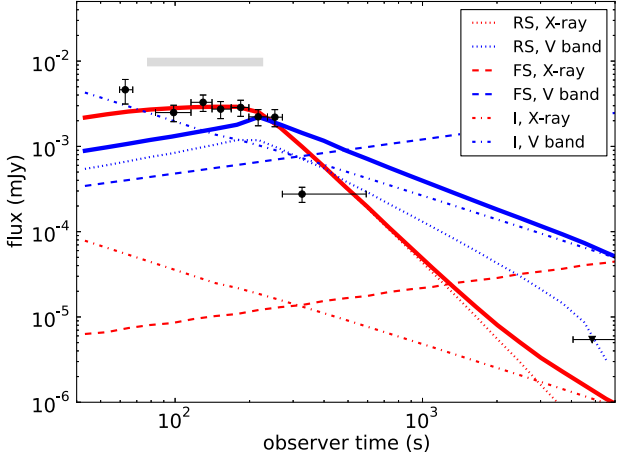}
\caption{X-ray and R-band light curves for GRB 120521A. The black data points are \emph{Swift} XRT measurements at 1 keV, the horizontal grey bar indicates an optical upper limit. The red curves represent model-based X-rays, the blue curves optical flux. The RS shock contributions are shown by dotted lines, the plateau FS contribution by a dashed lines and the impulsive FS emission by dash-dotted line. The resulting combined fluxes (solid lines) are obtained by adding the RS contribution to the FS contribution, switching between plateau and injected FS when these cross each other. The RS emission is switched off in the lab frame of the explosion once the last injected energy has passed through the RS. The model parameters are $z = 0.4$, $d_L = 6.3 \times 10^{27}$ cm, $E_{iso} = 9 \times 10^{50}$ erg, $T_{in} = 180$ s, $\epsilon_e = 5.6 \times 10^{-2}$, $\epsilon_{B, RS} = 7.9 \times 10^{-3}$, $\epsilon_{B, FS} = 3.2 \times 10^{-6}$, $\xi_N = 1$, $n_{ref} = 2$ cm$^{-3}$, $\eta = 1.1 \times 10^5$, $p = 2.2$, $k = 0$, $q = 0$. The actual redshift for this burst is not known.}
 \label{GRB120521A_figure}
\end{figure}

\begin{figure}
 \centering
  \includegraphics[width=\columnwidth]{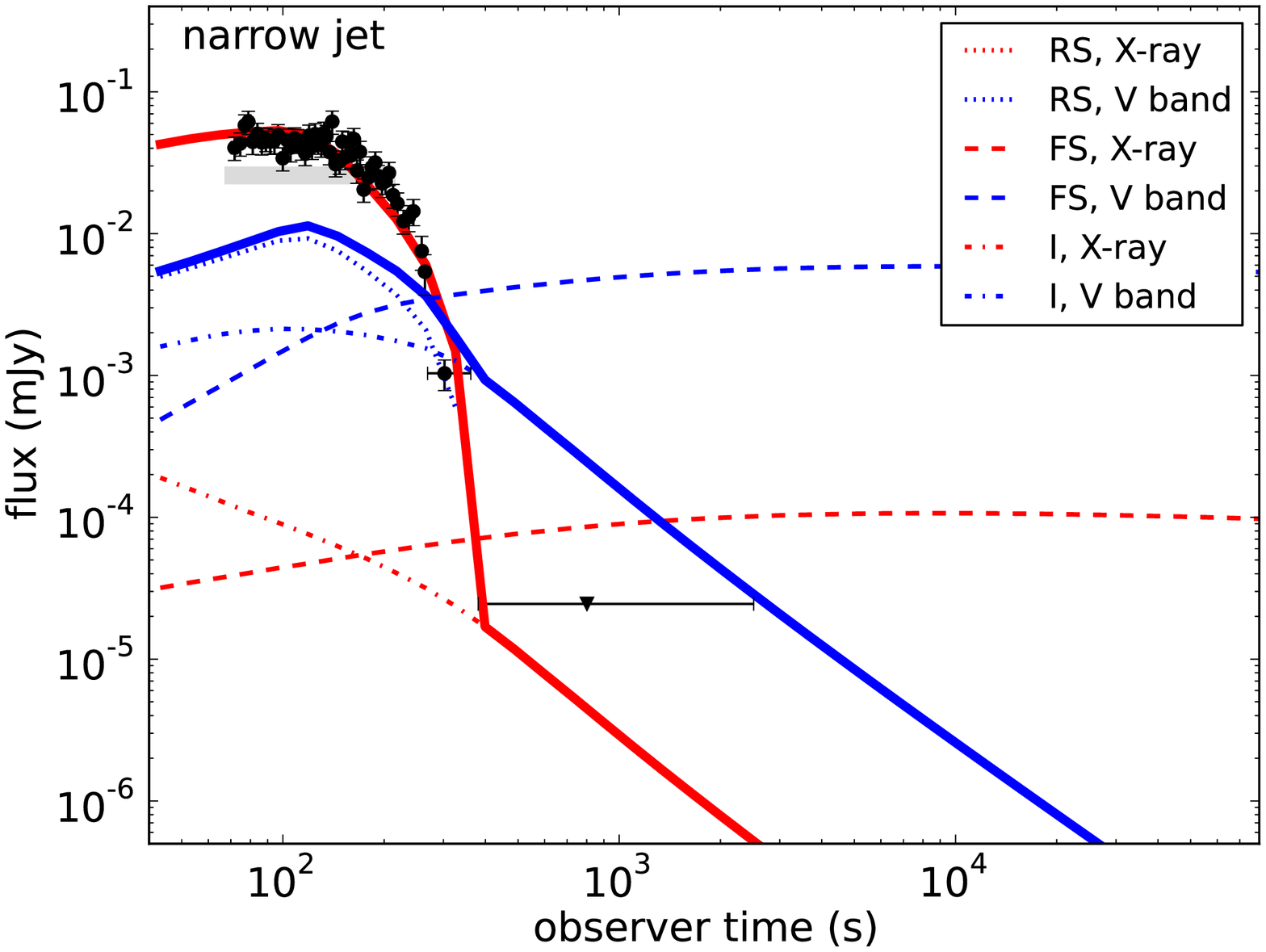}
  \includegraphics[width=\columnwidth]{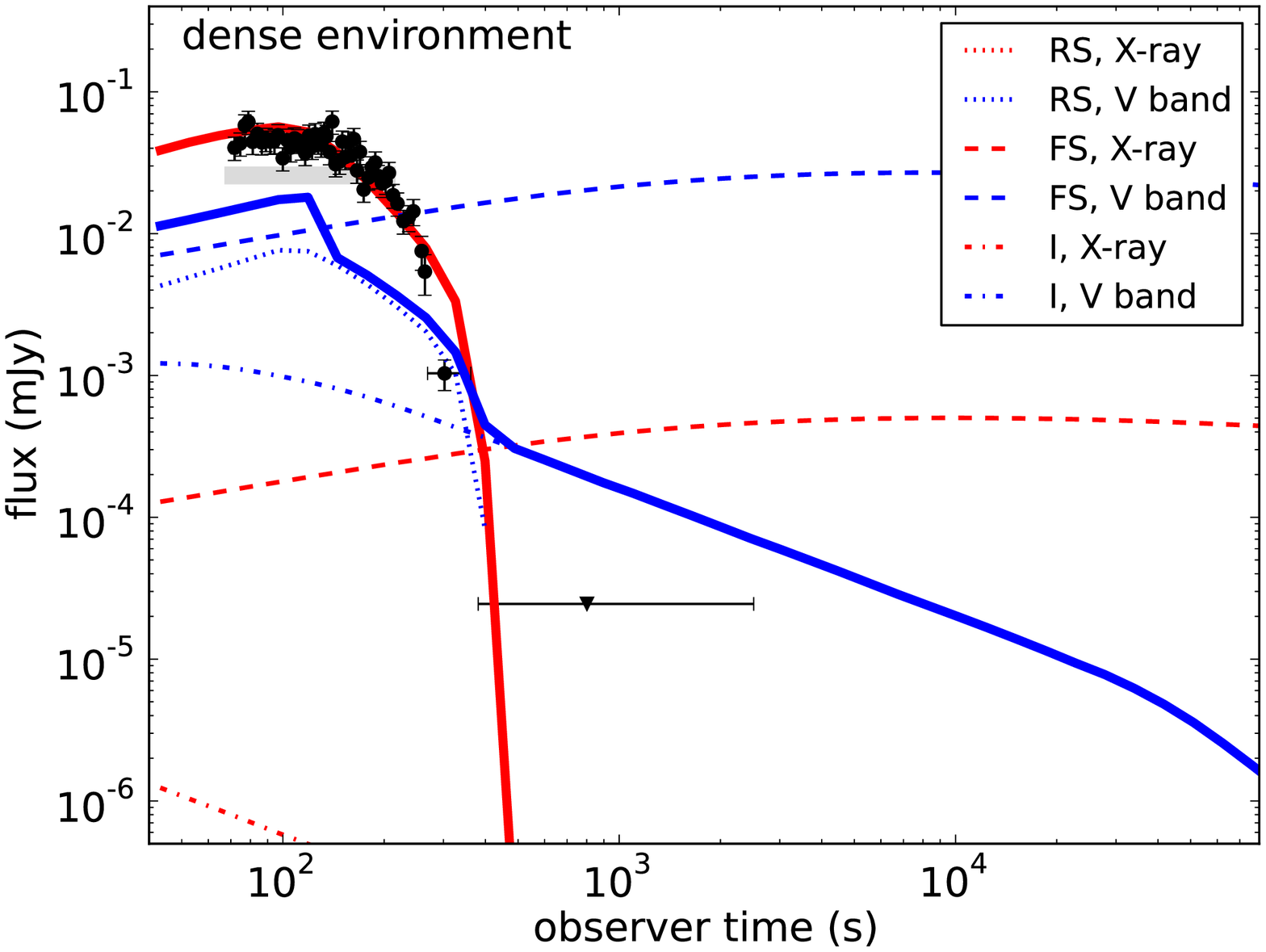}
\caption{Same as Fig. \ref{GRB120521A_figure}, now for GRB 090515. For the top plot, the model parameters are $z = 0.4$, $d_L = 6.3 \times 10^{27}$ cm, $E_{iso} = 1.2 \times 10^{51}$ erg, $T_{in} = 10^2$ s, $\epsilon_e = 0.5$, $\epsilon_{B, RS} = 3.9 \times 10^{-2}$, $\epsilon_{B, FS} = 10^{-6}$, $\xi_N = 0.79$, $n_{ref} = 0.16$ cm$^{-3}$, $\eta = 2.4 \times 10^4$, $p = 2.2$, $k = 0$, $q = 0$, $\theta_0 = 0.86^\circ$. For the bottom plot, the differing model parameters are $E_{iso} = 2.1 \times 10^{51}$ erg , $T_{in} = 2 \times 10^2$ s., $\epsilon_{B, FS} = 4 \times 10^{-6}$, $\epsilon_{B, RS} = 10^{-9}$, $\xi_N = 1$, $n_{ref} = 1 \times 10^5$ cm$^{-3}$, $\eta = 1.2 \times 10^4$, $\theta_0 = 5^\circ$, with $n_{ref}$ dropping to $0.1$ cm$^{-3}$ at radii beyond $\sim 6 \times 10^{14}$ cm, associated with the arrival of the last injected energy. These values do \emph{not} represent the best possible fit, but demonstrate a proof of principle. The actual redshift for this burst is not known.}
 \label{GRB090515_figure}
\end{figure}

In \cite{Leventis2014}, it is demonstrated how the prolonged energy injection scenario can be used to explain observations of regular afterglow plateaus, using GRB's 080928 and 090423. To further test the range of the applicability of this type of model we now turn to two more extreme cases, GRB120521A and 090515, characterized by an inferred strong post-plateau decline. A number of recent papers explain short GRB plateaus from a magnetar model \citep{Rowlinson2010, Rowlinson2013, Gompertz2013}, and it is instructive to see if these light curves can in principle be accounted for by a RS-FS system, while remaining agnostic about the nature of the power source (it can also be a magnetar-driven FS-RS system, which are argued to have $q = 0$, \citealt{Dai1998, ZhangMeszaros2001}).

Fig.  \ref{GRB120521A_figure} shows a comparison between BM solution-based light curves and \emph{Swift} XRT data and optical upper limits for GRB 120521A. The synthetic light curves were obtained using the analytical solution described in section \ref{self-similar_section} in combination with the synchrotron radiation module from \cite{vanEerten2009, vanEerten2010transrelativistic} and the synchrotron prescription from \cite{vanEerten2010offaxis} (as was done at the start of this section). Once the last injected energy passes through the RS, the RS emission is turned off in the lab frame, leading to post-plateau slopes dictated by the curvature effect. In this case I assumed spherical emission, but the flux would be identical for collimated flow unless the jet were very narrow ($\theta_0 < \sim 1^{\circ}$)

The figure illustrates that it is indeed possible to describe the data using plausible model parameters, if one accepts a low magnetic field for the FS (low being a relative term, these values are fully consistent with e.g. \citealt{Kumar2010, Santana2013}). The emission from this burst was argued to require a magnetar origin (as opposed to synchrotron emission from a blast wave) by \cite{Rowlinson2013} based on their inferred steep post-plateau slope and assumed difficulty to account for the optical upper limit from a synchrotron spectrum. Because the inferred steep plateau for this burst is based mainly on a single data point with large error in time, results consistent with the data can also be achieved without equally steep slope. The optical upper limit was considered inconsistent with a synchrotron model based on assuming at most the presence of the cooling break $\nu_c$ between optical and X-rays. In our case the issue is avoided by high values for $\nu_m$. As discussed previously, this is naturally expected when RS emission dominates in the plateau phase. For a smooth spectral break, it is not even necessary that $\nu_m > \nu$ in the optical.

The case of GRB 090515, shown in Fig. \ref{GRB090515_figure}, is more challenging, even though the inferred post-plateau slope is far less extreme than that of GRB 120521A. But here the post-plateau slope is not set by a single data point (although it requires a re-binning of the automatically generated light curve presented on the XRT website to bring this aspect to the surface; for 120521A, the photon arrival times are spread out to much for re-binning to make a difference). The figure presents two alternatives for qualitatively reproducing the features of this burst. In both cases it is necessary to circumvent the limitation on the maximum steepness of the post-plateau slope imposed by the curvature effect (i.e. the spread in arrival times of emission from different angles, even when RS emission ceases at a single lab frame time). Since the curvature effect only applies as long as the jet nature of the outflow is not apparent to the observer, it is necessary that the burst is observed at a time when the blast wave has decelerated at least beyond the point where $\gamma \sim 1 / \theta_0$ (see section \ref{sideways_spreading_subsection} below and e.g. \citealt{Granot2007, vanEerten2013proceedings} for extensive discussions of the nature of jet breaks). In other words, either $\gamma$ or $\theta_0$ has to be such that this is the case already around $10^2$ s.

The top plot of fig. \ref{GRB120521A_figure} uses a very narrow jet with $\theta_0 = 0.9^{\circ}$. On the one hand, narrow jets of a few degrees (e.g. five degrees or lower) have actually been inferred for multiple short GRBs \citep{Stratta2007, NicuesaGuelbenzu2012, Fong2012, Berger2013}. On the other hand, it is not easy to come up with a plausible mechanism to create a narrow jet from a neutron star merger, the preferred scenario for short GRBs.

Short GRBs are often found at an offset from their host galaxies (see e.g. \citealt{Fong2010}), providing a natural explanation for the low circumburst densities that are often inferred from afterglow modeling (e.g. \citealt{Belczynski2006, Berger2013}). It should however be kept in mind that, on the whole, short GRB circumburst densities are poorly constrained due to lack of full broadband coverage (that would need to cover all spectral regimes in order to fully constrain the model). Also, the distance $\sim 10^{13-14}$ cm, covered by the blast wave during the short GRB plateau phase is still very close to the progenitor system and therefore sensitive to its history. When probed at later times and further distant from the source (e.g. $\sim 10^{15}$ cm for GRB 050724, \citealt{Berger2005}), the local density will more likely resemble the environment density more closely and the effect of density perturbations closer to the source will be negligible compared to the total integrated density that shapes the blast wave evolution. In the bottom plot of fig. \ref{GRB120521A_figure}, I fixed the opening angle of the jet to $5^\circ$. The early jet break is now achieved with a circumburst density out to $c T_{last}$ of $n_{ref} = 10^5$ cm$^{-3}$. Integrated over radius, this is about $10^{-5} M_\odot$. Since the Lorentz factor of the outflow is determined by the total amount of swept-up mass, the same effect can be reached by having more circumburst mass within a smaller radius. The total amount of mass remains tiny compared to the $\sim 0.1 M_\odot$ expected to be ejected closely before the merger occurs (see e.g. \citealt{Rosswog2013}). Once the blast wave pierces the massive shell and emerges in the dilute environment, the FS is expected to speed up essentially instantaneously to a new Lorentz factor dictated by the ratio between the two densities \citep{Gat2013}.

\section{Additional discussion}
\label{discussion_section}

\subsection{Charged particle acceleration and emission}
\label{cooling_subsection}

The detailed physics of charged particle acceleration and magnetic field generation in relativistic blast waves and turbulent flows are an incredibly complicated subject that is still poorly understood, and far beyond the scope of this paper. Instead, I have used the (commonly applied) simplifying parameters $\epsilon_B$, $\epsilon_e$ and $\xi_N$, along with some implicit assumptions about their downstream evolution (see \citealt{vanEerten2013proceedings} for details). Furthermore, a global approach was used to obtain a single cooling time for the entire plasma, rather than accounting for changing cooling times for fluid parcels as they advect away from acceleration sites. A more detailed treatment will shift the position of $\nu_c$ within the RS and FS spectra (see also \citealt{vanEerten2010offaxis}). Such a treatment, however, would need to address the question at which place(s) electrons are accelerated into a non-thermal distribution. Maybe the dominant process is Fermi shock acceleration at the FS and RS shock fronts. Local cooling could then, in principle, be calculated within the self-similar BM framework using the advection equation \ref{soundwave_equation} from section \ref{transition_section}. However, a (relativistic) FS-CD-RS system can easily be Rayleigh-Taylor unstable at the CD (see e.g. \citealt{Duffell2013} for a recent demonstration). Particle acceleration can then also take place throughout the turbulent region, which will spread out through the RS and FS regions.

\subsection{sideways spreading}
\label{sideways_spreading_subsection}

Once causal contact is established among all angles of the collimated blast wave, deviations from radial flow can be expected. This sideways spreading is partially responsible for what is observed as the `jet break' in the afterglow light curve (the other part being the edges of the outflow becoming visible). The onset of spreading can be calculated as the arrival time of a relativistic sound wave moving from edge to tip along a shock front decelerating according to $\Gamma \propto t^{-m/2}$. For nonzero $m$, this yields $\theta_0 = 1 / [(3-k) \Gamma]$ \citep{vanEerten2013proceedings, MacFadyen2013}. For our typical ISM scenario with $q = 0$ and opening angle 0.1, this leads to $\Gamma = 10$, which puts the onset of spreading well after $\Gamma = 26.8$, when the last energy has been delivered to the shock front. In the typical wind scenario, $m = 0$, so we can not use $\Gamma$ as a measure of the amount of time that has passed. In this case, $t = t_0 \exp[4 \theta_0 \Gamma]$, for a sound wave departing from the edge at $t_0$. Since $t_0 \sim R_0 / c$, causal contact along the shock front is therefore expected already early on during the injection phase.

Another way of assessing the importance of sideways spreading is by comparing injected energy and sideways energy loss, in terms of luminosities $L_{IN}$, $L_{OUT}$, respectively. Using the time derivative of the collimation-corrected amount of energy within the blast wave (approximated as a cylinder with homogeneous pressure set by the shock-jump conditions and width set by $\chi = 1$, $\chi_{RS}$) and sideways energy flux $F = \sqrt{2} \Gamma p c \beta_\perp'$ (where $\beta_\perp'$ the sideways velocity in the radially comoving frame, somewhere between sound speed $1 / \sqrt{3}$ and light speed 1), we obtain
\begin{equation}
\frac{L_{IN}}{L_{OUT}} = \frac{\alpha (3 -k - m) \sqrt{2}}{\chi_{RS} \beta_\perp'} \theta_0 \Gamma,
\end{equation}
where $\alpha$ a term of order unity whose definition can be found in BM76. Since the other terms are also of order unity, the balance between energy flows is essentially set by $\theta_0 \Gamma$. The result is therefore similar to what is inferred from looking at sound waves along the shock front. For constant injection with $q = 0$, ISM blast waves remain radial until the jet break. Constant injection wind blast waves remain radial throughout the energy injection stage, except for a minor bow shock (if $\Gamma > \theta_0$), or start spreading immediately until $\theta \sim \Gamma^{-1}$, which effectively puts a lower limit on how narrowly energy injection can be collimated when $m = 0$ (as is the case for a wind with $q=0$, $k =2$; note that the narrow jet treatment of GRB 090515 from the preceding section was done in the context of an ISM-type environment where $m=1$).

In a practical sense, this implies that the results obtained in this paper, both analytical and numerical, are approximately applicable even for blast waves that spread out at a later stage and no two-dimensional energy injection simulations are required. Simulation-based fit codes based on templates for impulsive energy injection blast waves \citep{vanEerten2012boxfit, vanEerten2012scalings} can be extended to include an energy injection stage by attaching a series of one-dimensional simulation-based templates for various injection durations. As discussed previously in section \ref{light_curve_section}, it is straightforward to extend scale invariance to include this stage. 

What this does \emph{not} mean, however, is that no jet break-like effect can be seen in the light curve during energy ejection. For nonzero $m$, there remains the effect of increasing relativistic beaming cone width even while the outflow remains radial. Given that afterglows are typically observed off-axis \citep{Ryan2013, Ryan2014}, it is even likely that the onset of this type of jet break even occurs during the injection phase of the blast wave.

\subsection{Energy injection by massive ejecta}

There exists an extensive literature on afterglow emission from massive ejecta (e.g. \citealt{Sari1995, Kobayashi1999, Kobayashi2000, RamirezRuiz2002, Wu2003, Peng2005, Zou2005, Yi2013, Leventis2014}). Prolonged energy injection can be understood as a generalization of massive ejecta. In the latter case, the cold ejecta will stratify into a cold wind-type outflow (see also section \ref{free_flowing_wind_subsection}) of width $\Delta R \propto R / \Gamma^2$. As long as the RS has not fully crossed the ejecta, this situation is identical to a constant energy input (in the form of ejecta kinetic energy) with $q = 0$ (and in the `thick-shell' case, where the RS becomes relativistic in the frame of the ejecta before completing its crossing, the self-similar profiles described in this paper will also arise). Therefore, the flux equations derived in this paper will reduce to those presented e.g. by \cite{Yi2013} if one takes $q = 0$. For ultrarelativistic ejecta, characteristic moments in the ejecta evolution like the completion of the reverse shock crossing and the deceleration radius are expected to occur in the observer frame on a timescale far smaller than the typically observed plateau end times of $10^{3-4}$ s. Less relativistic massive ejecta, such as the cocoon surrounding the collapsar jet, will have comparable timescales to the observed plateau durations. In that case the RS is likely to remain non-relativistic, so the full self-similar BM profile will not emerge. Nevertheless, equating the end-time of the plateau to the deceleration time and measuring the flux at this time will pick out the same point in the characteristic evolution across bursts, in the same way as measuring the flux at $T_{in}$ does.

\subsection{The Lorentz factors of the outflow}
\label{Lorentz_factor_section}

Right now, the solutions presented in this work are based on a number of (connected) assumptions: the free-flowing relativistic wind is ultra-relativistic with Lorentz factor significantly exceeding that of the FS; The engine activity is long-lived, allowing for a relativistic RS to emerge; The Lorentz factor of the free-flowing wind remains fixed at $\eta$, restricting the applicability of the solution to radii $> \eta R_0$ and requiring an unchanging ratio between mass loss and luminosity of the engine. It should be noted that there are other ways of generalizing the standard fireball model of instantaneous massive ejecta, that do not necessarily lead to a relativistic RS. It is possible to relax the constraint that $\eta$ remain fixed. If $\eta$ changes over time, but the wind Lorentz factor at the RS remains larger than the FS Lorentz factor, the BM solution described in this paper still applies. The only difference then becomes that the time-dependence of $\eta$ needs to be carried over to the density profile. If $\eta \propto t_{in}^s$, this can be implemented in the flux equations by replacing $\eta \to \eta_{ref} (t_{obs} / t_{ref})^s$ (ignoring redshift effects; once again the factors $\Gamma^2$ cancel in going from emission at the source to arrival at RS and from there to observer time). Relaxing the requirement that the engine remains active for a long time, quickly leads to a situation where the RS is not relativistic. At this point, the flux equations for the RS region (but not the FS region), quickly become qualitatively different. Under specific assumptions for the density profile of the ejecta and its acceleration behind the RS, self-similar solutions with Newtonian RS remain possible. An example can be found in the paper by \cite{Nakamura2006} mentioned in the introduction. The flux equations for non-relativistic RS systems in general density profiles are provided by \cite{Yi2013}.

\section{Summary}
\label{summary_section}

The self-similar forward-shock-reverse-shock (FS-RS) profile arising from a powerful astrophysical source, with long-term luminosity depending on time according to a power law, is studied in detail. A treatment of the density profile and evolution in the reverse shock region is added to the self-similar solution for the fluid profile from \cite{Blandford1976}, which is valid as long as the Lorentz factor of the relativistic wind carrying the injected energy greatly exceeds that of the forward shock. The ratio of downstream densities behind the RS and FS remains fixed for constant energy injection into a wind-type environment, and decreases linearly for constant injection into a homogeneous environment. For typical long GRB afterglow parameters, the FS Lorentz factor $\Gamma$ will be around 20-30 by the end of the plateau stage around $10^4$ s, with the Lorentz factor in the wind case having remained unchanged over time. The self-similar fluid profile obeys the same scale invariances as in the impulsive energy injection case \citep{vanEerten2012boxfit, vanEerten2012scalings}, although the Lorentz factor $\eta$ of the inflowing ultra-relativistic wind and the coefficient of the power-law luminosity $q$ increase the dimensionality of the parameter space. 

The self-similar profile is confirmed by high-resolution numerical simulations in one dimension of power law injection onto a computational grid. The simulations also confirm that the transition to an impulsive energy injection profile, following the cessation of energy injection, takes about a sound-crossing time. This crossing time can be calculated exactly from the analytical solution. The assumption of radial flow within a cone of angle $\theta_0$ remains valid up until the jet break in case the forward shock Lorentz factor decreases over time (e.g. ISM-type environments with constant injection) and requires energy injection angle $\theta_0 \gg \Gamma$ if $\Gamma$ remains constant (e.g. constant injection in the wind case). In the latter case, if $\theta_0 < \Gamma$, the jet initially spreads quickly and an effective energy injection angle $\theta \sim \Gamma^{-1}$ is maintained.

When combined with a standard synchrotron approach to radiation from shock-accelerated electrons (e.g. \citealt{Sari1998}), the resulting model dependencies of the flux equations for all spectral regimes of FS and RS emission, as well as the flux levels, can be calculated. These are provided in tables. The flux equations and an application using `typical' afterglow parameters reveal the following properties of the combined emission from these systems:

\begin{itemize}
  \item The contribution from the reverse shock region can easily be significant or even dominant, certainly when different magnetizations for both regions are taken into account.
  \item The observed light curve evolution will show a complex interplay between changing FS and RS contributions
  \item For limited injection duration, the transition from a light curved shaped by sustained injection to one shaped by impulsive energy injection occurs around the point when the cessation of energy injection is communicated to the shock front, although differences in arrival time of emission from different angles will spread out this feature over time.
  \item Because the synchrotron break frequency $\nu_m$ depends quadratically on $\eta$ for RS emission, it tends to be shifted to high frequencies and can be seen crossing the optical bands during the plateau phase for certain combinations of model parameters.
  \item For constant energy injection, the RS emission for a blast wave moving into a wind environment tends to be in the fast cooling regime during the plateau stage, while RS emission in the homogeneous case tends to be in the slow cooling regime. Conclusions about the regime based on post-plateau FS emission can not be extrapolated back into the plateau stage when RS emission is dominant at that time. In addition, for a power law distribution of accelerated particles with slope $-p \sim -2.2$, the spectra for both regimes are close to identical.
  \item The observational optical and X-ray $F_b$ - $t_b$ correlations between flux at the end of the plateau and the plateau duration follow naturally from RS dominated emission and values consistent within $2 \sigma$ can be obtained for FS dominated emission. The wind scenario leads to values marginally closer to the prediction, but the difference is very small. The correlation emerges even across bursts with otherwise differing values for model parameters ($E_{iso}$, $\epsilon_B$, $n_0$, etc.), as these differences will lead to scatter in the correlation, but not impact its slope. The correlations are not affected by the value of $q$, but do require $E_{iso}$ and $T_{in}$ to be independent. A comparison between the correlation predictions and model fluxes in a more realistic setting, using population studies of synthetic afterglows, is the topic of a follow-up paper \citep{vanEerten2014correlations}.
\end{itemize}

Finally, it is shown that the model of synchrotron emission from a blast wave with sustained energy injection up to $\sim 10^2$ seconds can in principle be used to explain short GRBs 090515 and 120521A, that were previously argued to be inconsistent with a synchrotron model and require radiation directly from a magnetar instead \citep{Rowlinson2013}. In our demonstration, a power law slope $q = 0$ was shown consistent with the data. This is actually consistent with energy injection from a magnetar into a FS-RS system, but the generic power law luminosity assumption allows one to remain agnostic about whether or not 090515 and 120521A were caused by magnetars. The sudden cessation of energy injection required in the FS-RS explanation can be understood from a magnetar source. Or it might result from complex fluid behavior once the power law for the luminosity drops below -1, which becomes asymptotically equivalent to impulsive energy injection. The profile at the back end of the injected material can be steepened by rarefaction waves, depending on the shape of the transition.

\section*{Acknowledgments}

I wish to thank Antonia Rowlinson and Patricia Schady for assisting with \emph{Swift} data and Kostadinos Leventis, Alexander van der Horst, Paul Duffell and Andrew MacFadyen for helpful discussion. This research was supported in part through Chandra grant TM3-14005X and by the Alexander von Humboldt foundation. I gratefully acknowledge the referee Ehud Nakar for his corrections and suggested revisions.

\bibliographystyle{mn2e}
\bibliography{selfsimilarBM}

\begin{thebibliography}{102}
\expandafter\ifx\csname natexlab\endcsname\relax\def\natexlab#1{#1}\fi

\bibitem[{{Belczynski} {et~al}\mbox{.}(2006){Belczynski}, {Perna}, {Bulik},
  {Kalogera}, {Ivanova}, \& {Lamb}}]{Belczynski2006}
{Belczynski} K., {Perna} R., {Bulik} T., {Kalogera} V., {Ivanova} N., {Lamb}
  D.~Q., 2006, \apj, 648, 1110

\bibitem[{{Beloborodov} \& {Uhm}(2006)}]{Beloborodov2006}
{Beloborodov} A.~M., {Uhm} Z.~L., 2006, \apjl, 651, L1

\bibitem[{{Berger}(2013)}]{Berger2013}
{Berger} E., 2013, ArXiv e-prints

\bibitem[{{Berger} {et~al}\mbox{.}(2005){Berger}, {Price}, {Cenko}, {Gal-Yam},
  {Soderberg}, {Kasliwal}, {Leonard}, {Cameron}, {Frail}, {Kulkarni}, {Murphy},
  {Krzeminski}, {Piran}, {Lee}, {Roth}, {Moon}, {Fox}, {Harrison}, {Persson},
  {Schmidt}, {Penprase}, {Rich}, {Peterson}, \& {Cowie}}]{Berger2005}
{Berger} E. {et~al.}, 2005, \nat, 438, 988

\bibitem[{{Blandford} \& {McKee}(1976)}]{Blandford1976}
{Blandford} R.~D., {McKee} C.~F., 1976, Physics of Fluids, 19, 1130

\bibitem[{{Blandford} \& {McKee}(1977)}]{Blandford1977}
{Blandford} R.~D., {McKee} C.~F., 1977, \mnras, 180, 343

\bibitem[{{Butler} \& {Kocevski}(2007)}]{ButlerKocevski2007}
{Butler} N.~R., {Kocevski} D., 2007, \apj, 668, 400

\bibitem[{{Chevalier} \& {Li}(2000)}]{Chevalier2000}
{Chevalier} R.~A., {Li} Z.-Y., 2000, \apj, 536, 195

\bibitem[{{Curran} {et~al}\mbox{.}(2009){Curran}, {Starling}, {van der Horst},
  \& {Wijers}}]{Curran2009}
{Curran} P.~A., {Starling} R.~L.~C., {van der Horst} A.~J., {Wijers}
  R.~A.~M.~J., 2009, \mnras, 395, 580

\bibitem[{{Dai} \& {Lu}(1998)}]{Dai1998}
{Dai} Z.~G., {Lu} T., 1998, \aap, 333, L87

\bibitem[{{Dainotti}, {Cardone} \& {Capozziello}(2008){Dainotti}, {Cardone}, \&
  {Capozziello}}]{Dainotti2008}
{Dainotti} M.~G., {Cardone} V.~F., {Capozziello} S., 2008, \mnras, 391, L79

\bibitem[{{Dainotti} {et~al}\mbox{.}(2013){Dainotti}, {Petrosian}, {Singal}, \&
  {Ostrowski}}]{Dainotti2013}
{Dainotti} M.~G., {Petrosian} V., {Singal} J., {Ostrowski} M., 2013, \apj, 774,
  157

\bibitem[{{Dainotti} {et~al}\mbox{.}(2010){Dainotti}, {Willingale},
  {Capozziello}, {Fabrizio Cardone}, \& {Ostrowski}}]{Dainotti2010}
{Dainotti} M.~G., {Willingale} R., {Capozziello} S., {Fabrizio Cardone} V.,
  {Ostrowski} M., 2010, \apjl, 722, L215

\bibitem[{{Duffell} \& {MacFadyen}(2013)}]{Duffell2013}
{Duffell} P.~C., {MacFadyen} A.~I., 2013, \apj, 775, 87

\bibitem[{{Eichler} {et~al}\mbox{.}(1989){Eichler}, {Livio}, {Piran}, \&
  {Schramm}}]{Eichler1989}
{Eichler} D., {Livio} M., {Piran} T., {Schramm} D.~N., 1989, \nat, 340, 126

\bibitem[{{Evans}(2009)}]{Evans2009}
{Evans} P.~A. e.~a., 2009, \mnras, 397, 1177

\bibitem[{{Fong}, {Berger} \& {Fox}(2010){Fong}, {Berger}, \& {Fox}}]{Fong2010}
{Fong} W., {Berger} E., {Fox} D.~B., 2010, \apj, 708, 9

\bibitem[{{Fong} {et~al}\mbox{.}(2012){Fong}, {Berger}, {Margutti}, {Zauderer},
  {Troja}, {Czekala}, {Chornock}, {Gehrels}, {Sakamoto}, {Fox}, \&
  {Podsiadlowski}}]{Fong2012}
{Fong} W. {et~al.}, 2012, \apj, 756, 189

\bibitem[{{Fryxell} {et~al}\mbox{.}(2000){Fryxell}, {Olson}, {Ricker},
  {Timmes}, {Zingale}, {Lamb}, {MacNeice}, {Rosner}, {Truran}, \&
  {Tufo}}]{Fryxell2000}
{Fryxell} B. {et~al.}, 2000, \apjs, 131, 273

\bibitem[{{Gat}, {van Eerten} \& {MacFadyen}(2013){Gat}, {van Eerten}, \&
  {MacFadyen}}]{Gat2013}
{Gat} I., {van Eerten} H., {MacFadyen} A., 2013, \apj, 773, 2

\bibitem[{{Gehrels} {et~al}\mbox{.}(2004){Gehrels}, {Chincarini}, {Giommi},
  {Mason}, {Nousek}, {Wells}, {White}, {Barthelmy}, {Burrows}, {Cominsky},
  {Hurley}, {Marshall}, {M{\'e}sz{\'a}ros}, {Roming}, {Angelini}, {Barbier},
  {Belloni}, {Campana}, {Caraveo}, {Chester}, {Citterio}, {Cline}, {Cropper},
  {Cummings}, {Dean}, {Feigelson}, {Fenimore}, {Frail}, {Fruchter}, {Garmire},
  {Gendreau}, {Ghisellini}, {Greiner}, {Hill}, {Hunsberger}, {Krimm},
  {Kulkarni}, {Kumar}, {Lebrun}, {Lloyd-Ronning}, {Markwardt}, {Mattson},
  {Mushotzky}, {Norris}, {Osborne}, {Paczynski}, {Palmer}, {Park}, {Parsons},
  {Paul}, {Rees}, {Reynolds}, {Rhoads}, {Sasseen}, {Schaefer}, {Short},
  {Smale}, {Smith}, {Stella}, {Tagliaferri}, {Takahashi}, {Tashiro},
  {Townsley}, {Tueller}, {Turner}, {Vietri}, {Voges}, {Ward}, {Willingale},
  {Zerbi}, \& {Zhang}}]{Gehrels2004}
{Gehrels} N. {et~al.}, 2004, \apj, 611, 1005

\bibitem[{{Giannios}, {Mimica} \& {Aloy}(2008){Giannios}, {Mimica}, \&
  {Aloy}}]{Giannios2008}
{Giannios} D., {Mimica} P., {Aloy} M.~A., 2008, \aap, 478, 747

\bibitem[{{Gompertz} {et~al}\mbox{.}(2013){Gompertz}, {O'Brien}, {Wynn}, \&
  {Rowlinson}}]{Gompertz2013}
{Gompertz} B.~P., {O'Brien} P.~T., {Wynn} G.~A., {Rowlinson} A., 2013, \mnras,
  431, 1745

\bibitem[{{Goodman}(1986)}]{Goodman1986}
{Goodman} J., 1986, \apjl, 308, L47

\bibitem[{{Granot}(2007)}]{Granot2007}
{Granot} J., 2007, in Revista Mexicana de Astronomia y Astrofisica Conference
  Series, Vol.~27, Revista Mexicana de Astronomia y Astrofisica, vol. 27, pp.
  140--165

\bibitem[{{Granot}, {Piran} \& {Sari}(1999){Granot}, {Piran}, \&
  {Sari}}]{Granot1999}
{Granot} J., {Piran} T., {Sari} R., 1999, \apj, 513, 679

\bibitem[{{Granot} \& {Sari}(2002)}]{Granot2002}
{Granot} J., {Sari} R., 2002, \apj, 568, 820

\bibitem[{{Grupe} {et~al}\mbox{.}(2013){Grupe}, {Nousek}, {Veres}, {Zhang}, \&
  {Gehrels}}]{Grupe2013}
{Grupe} D., {Nousek} J.~A., {Veres} P., {Zhang} B.-B., {Gehrels} N., 2013,
  \apjs, 209, 20

\bibitem[{{Gruzinov} \& {Waxman}(1999)}]{Gruzinov1999}
{Gruzinov} A., {Waxman} E., 1999, \apj, 511, 852

\bibitem[{{Kobayashi}, {Piran} \& {Sari}(1999){Kobayashi}, {Piran}, \&
  {Sari}}]{Kobayashi1999}
{Kobayashi} S., {Piran} T., {Sari} R., 1999, \apj, 513, 669

\bibitem[{{Kobayashi} \& {Sari}(2000)}]{Kobayashi2000}
{Kobayashi} S., {Sari} R., 2000, \apj, 542, 819

\bibitem[{{Kumar} \& {Barniol Duran}(2010)}]{Kumar2010}
{Kumar} P., {Barniol Duran} R., 2010, \mnras, 409, 226

\bibitem[{{Kumar} \& {Panaitescu}(2000)}]{KumarPanaitescu2000}
{Kumar} P., {Panaitescu} A., 2000, \apjl, 541, L51

\bibitem[{{Leventis}, {Wijers} \& {van der Horst}(2014){Leventis}, {Wijers}, \&
  {van der Horst}}]{Leventis2014}
{Leventis} K., {Wijers} R.~A.~M.~J., {van der Horst} A.~J., 2014, \mnras, 437,
  2448

\bibitem[{{Li} {et~al}\mbox{.}(2012){Li}, {Liang}, {Tang}, {Chen}, {Xi},
  {L{\"u}}, {Gao}, {Zhang}, {Zhang}, {Yi}, {Lu}, {L{\"u}}, \& {Wei}}]{Li2012}
{Li} L. {et~al.}, 2012, \apj, 758, 27

\bibitem[{{Liang} {et~al}\mbox{.}(2008){Liang}, {Racusin}, {Zhang}, {Zhang}, \&
  {Burrows}}]{Liang2008}
{Liang} E.-W., {Racusin} J.~L., {Zhang} B., {Zhang} B.-B., {Burrows} D.~N.,
  2008, \apj, 675, 528

\bibitem[{{Lyutikov} \& {Blandford}(2003)}]{LyutikovBlandford2003}
{Lyutikov} M., {Blandford} R., 2003, ArXiv Astrophysics e-prints: 0312347

\bibitem[{{MacFadyen} \& {van Eerten}(2014)}]{MacFadyen2013}
{MacFadyen} A.~I., {van Eerten} H.~J., 2014, manuscript in preparation

\bibitem[{{MacFadyen} \& {Woosley}(1999)}]{MacFadyen1999}
{MacFadyen} A.~I., {Woosley} S.~E., 1999, \apj, 524, 262

\bibitem[{{MacNeice} {et~al}\mbox{.}(2000){MacNeice}, {Olson}, {Mobarry}, {de
  Fainchtein}, \& {Packer}}]{MacNeice2000}
{MacNeice} P., {Olson} K.~M., {Mobarry} C., {de Fainchtein} R., {Packer} C.,
  2000, Computer Physics Communications, 126, 330

\bibitem[{{Margutti} {et~al}\mbox{.}(2013){Margutti}, {Zaninoni}, {Bernardini},
  {Chincarini}, {Pasotti}, {Guidorzi}, {Angelini}, {Burrows}, {Capalbi},
  {Evans}, {Gehrels}, {Kennea}, {Mangano}, {Moretti}, {Nousek}, {Osborne},
  {Page}, {Perri}, {Racusin}, {Romano}, {Sbarufatti}, {Stafford}, \&
  {Stamatikos}}]{Margutti2013}
{Margutti} R. {et~al.}, 2013, \mnras, 428, 729

\bibitem[{{M{\'e}sz{\'a}ros}(2006)}]{Meszaros2006}
{M{\'e}sz{\'a}ros} P., 2006, Reports on Progress in Physics, 69, 2259

\bibitem[{{Meszaros}, {Laguna} \& {Rees}(1993){Meszaros}, {Laguna}, \&
  {Rees}}]{Meszaros1993}
{Meszaros} P., {Laguna} P., {Rees} M.~J., 1993, \apj, 415, 181

\bibitem[{{Meszaros} \& {Rees}(1997{\natexlab{a}})}]{Meszaros1997}
{Meszaros} P., {Rees} M.~J., 1997{\natexlab{a}}, \apj, 476, 232

\bibitem[{{Meszaros} \& {Rees}(1997{\natexlab{b}})}]{MeszarosRees1997Poynting}
{Meszaros} P., {Rees} M.~J., 1997{\natexlab{b}}, \apjl, 482, L29

\bibitem[{{Meszaros}, {Rees} \& {Wijers}(1998){Meszaros}, {Rees}, \&
  {Wijers}}]{Meszaros1998}
{Meszaros} P., {Rees} M.~J., {Wijers} R.~A.~M.~J., 1998, \apj, 499, 301

\bibitem[{{Mimica}, {Giannios} \& {Aloy}(2009){Mimica}, {Giannios}, \&
  {Aloy}}]{Mimica2009}
{Mimica} P., {Giannios} D., {Aloy} M.~A., 2009, \aap, 494, 879

\bibitem[{{Nakamura} \& {Shigeyama}(2006)}]{Nakamura2006}
{Nakamura} K., {Shigeyama} T., 2006, \apj, 645, 431

\bibitem[{{Nakayama} \& {Shigeyama}(2005)}]{Nakayama2005}
{Nakayama} K., {Shigeyama} T., 2005, \apj, 627, 310

\bibitem[{{Nicuesa Guelbenzu} {et~al}\mbox{.}(2012){Nicuesa Guelbenzu},
  {Klose}, {Greiner}, {Kann}, {Kr{\"u}hler}, {Rossi}, {Schulze}, {Afonso},
  {Elliott}, {Filgas}, {Hartmann}, {K{\"u}pc{\"u} Yolda{\c s}}, {McBreen},
  {Nardini}, {Olivares E.}, {Rau}, {Schmidl}, {Schady}, {Sudilovsky}, {Updike},
  \& {Yolda{\c s}}}]{NicuesaGuelbenzu2012}
{Nicuesa Guelbenzu} A. {et~al.}, 2012, \aap, 548, A101

\bibitem[{{Nousek} {et~al}\mbox{.}(2006){Nousek}, {Kouveliotou}, {Grupe},
  {Page}, {Granot}, {Ramirez-Ruiz}, {Patel}, {Burrows}, {Mangano}, {Barthelmy},
  {Beardmore}, {Campana}, {Capalbi}, {Chincarini}, {Cusumano}, {Falcone},
  {Gehrels}, {Giommi}, {Goad}, {Godet}, {Hurkett}, {Kennea}, {Moretti},
  {O'Brien}, {Osborne}, {Romano}, {Tagliaferri}, \& {Wells}}]{Nousek2006}
{Nousek} J.~A. {et~al.}, 2006, \apj, 642, 389

\bibitem[{{Paczynski}(1986)}]{Paczynski1986}
{Paczynski} B., 1986, \apjl, 308, L43

\bibitem[{{Paczynski}(1991)}]{Paczynski1991}
{Paczynski} B., 1991, \actaa, 41, 257

\bibitem[{{Paczynski}(1998)}]{Paczynski1998}
{Paczynski} B., 1998, \apjl, 494, L45

\bibitem[{{Panaitescu}, {Meszaros} \& {Rees}(1998){Panaitescu}, {Meszaros}, \&
  {Rees}}]{Panaitescu1998}
{Panaitescu} A., {Meszaros} P., {Rees} M.~J., 1998, \apj, 503, 314

\bibitem[{{Panaitescu} \& {Vestrand}(2011)}]{PanaitescuVestrand2011}
{Panaitescu} A., {Vestrand} W.~T., 2011, \mnras, 414, 3537

\bibitem[{{Peng}, {K{\"o}nigl} \& {Granot}(2005){Peng}, {K{\"o}nigl}, \&
  {Granot}}]{Peng2005}
{Peng} F., {K{\"o}nigl} A., {Granot} J., 2005, \apj, 626, 966

\bibitem[{{Piran}(2004)}]{Piran2004}
{Piran} T., 2004, Reviews of Modern Physics, 76, 1143

\bibitem[{{Piran}, {Shemi} \& {Narayan}(1993){Piran}, {Shemi}, \&
  {Narayan}}]{Piran1993}
{Piran} T., {Shemi} A., {Narayan} R., 1993, \mnras, 263, 861

\bibitem[{{Racusin} {et~al}\mbox{.}(2009){Racusin}, {Liang}, {Burrows},
  {Falcone}, {Sakamoto}, {Zhang}, {Zhang}, {Evans}, \& {Osborne}}]{Racusin2009}
{Racusin} J.~L. {et~al.}, 2009, \apj, 698, 43

\bibitem[{{Racusin} {et~al}\mbox{.}(2011){Racusin}, {Oates}, {Schady},
  {Burrows}, {de Pasquale}, {Donato}, {Gehrels}, {Koch}, {McEnery}, {Piran},
  {Roming}, {Sakamoto}, {Swenson}, {Troja}, {Vasileiou}, {Virgili},
  {Wanderman}, \& {Zhang}}]{Racusin2011}
{Racusin} J.~L. {et~al.}, 2011, \apj, 738, 138

\bibitem[{{Ramirez-Ruiz}, {Celotti} \& {Rees}(2002){Ramirez-Ruiz}, {Celotti},
  \& {Rees}}]{RamirezRuiz2002}
{Ramirez-Ruiz} E., {Celotti} A., {Rees} M.~J., 2002, \mnras, 337, 1349

\bibitem[{{Rees} \& {Meszaros}(1998)}]{ReesMeszaros1998}
{Rees} M.~J., {Meszaros} P., 1998, \apjl, 496, L1

\bibitem[{{Rhoads}(1999)}]{Rhoads1999}
{Rhoads} J.~E., 1999, \apj, 525, 737

\bibitem[{{Rosswog}, {Piran} \& {Nakar}(2013){Rosswog}, {Piran}, \&
  {Nakar}}]{Rosswog2013}
{Rosswog} S., {Piran} T., {Nakar} E., 2013, \mnras, 430, 2585

\bibitem[{{Rowlinson} {et~al}\mbox{.}(2013){Rowlinson}, {O'Brien}, {Metzger},
  {Tanvir}, \& {Levan}}]{Rowlinson2013}
{Rowlinson} A., {O'Brien} P.~T., {Metzger} B.~D., {Tanvir} N.~R., {Levan}
  A.~J., 2013, \mnras, 430, 1061

\bibitem[{{Rowlinson} {et~al}\mbox{.}(2010){Rowlinson}, {O'Brien}, {Tanvir},
  {Zhang}, {Evans}, {Lyons}, {Levan}, {Willingale}, {Page}, {Onal}, {Burrows},
  {Beardmore}, {Ukwatta}, {Berger}, {Hjorth}, {Fruchter}, {Tunnicliffe}, {Fox},
  \& {Cucchiara}}]{Rowlinson2010}
{Rowlinson} A. {et~al.}, 2010, \mnras, 409, 531

\bibitem[{{Ryan}, {Van Eerten} \& {MacFadyen}(2013){Ryan}, {Van Eerten}, \&
  {MacFadyen}}]{Ryan2013}
{Ryan} G., {Van Eerten} H., {MacFadyen} A., 2013, 7th Huntsville Gamma-Ray
  Burst Symposium, GRB 2013: paper 30 in eConf Proceedings C1304143

\bibitem[{{Ryan}, {Van Eerten} \& {MacFadyen}(2014){Ryan}, {Van Eerten}, \&
  {MacFadyen}}]{Ryan2014}
{Ryan} G., {Van Eerten} H., {MacFadyen} A., 2014, Manuscript in preparation

\bibitem[{{Santana}, {Barniol Duran} \& {Kumar}(2013){Santana}, {Barniol
  Duran}, \& {Kumar}}]{Santana2013}
{Santana} R., {Barniol Duran} R., {Kumar} P., 2013, ArXiv e-prints: 1309.3277

\bibitem[{{Sari}(1997)}]{Sari1997}
{Sari} R., 1997, \apjl, 489, L37

\bibitem[{{Sari} \& {M{\'e}sz{\'a}ros}(2000)}]{SariMeszaros2000}
{Sari} R., {M{\'e}sz{\'a}ros} P., 2000, \apjl, 535, L33

\bibitem[{{Sari} \& {Piran}(1995)}]{Sari1995}
{Sari} R., {Piran} T., 1995, \apjl, 455, L143

\bibitem[{{Sari}, {Piran} \& {Narayan}(1998){Sari}, {Piran}, \&
  {Narayan}}]{Sari1998}
{Sari} R., {Piran} T., {Narayan} R., 1998, \apjl, 497, L17+

\bibitem[{{Shemi} \& {Piran}(1990)}]{Shemi1990}
{Shemi} A., {Piran} T., 1990, \apjl, 365, L55

\bibitem[{{Stratta} {et~al}\mbox{.}(2007){Stratta}, {D'Avanzo}, {Piranomonte},
  {Cutini}, {Preger}, {Perri}, {Conciatore}, {Covino}, {Stella}, {Guetta},
  {Marshall}, {Holland}, {Stamatikos}, {Guidorzi}, {Mangano}, {Antonelli},
  {Burrows}, {Campana}, {Capalbi}, {Chincarini}, {Cusumano}, {D'Elia}, {Evans},
  {Fiore}, {Fugazza}, {Giommi}, {Osborne}, {La Parola}, {Mineo}, {Moretti},
  {Page}, {Romano}, \& {Tagliaferri}}]{Stratta2007}
{Stratta} G. {et~al.}, 2007, \aap, 474, 827

\bibitem[{{Thompson}(1994)}]{Thompson1994}
{Thompson} C., 1994, \mnras, 270, 480

\bibitem[{{Troja} {et~al}\mbox{.}(2007){Troja}, {Cusumano}, {O'Brien}, {Zhang},
  {Sbarufatti}, {Mangano}, {Willingale}, {Chincarini}, {Osborne}, {Marshall},
  {Burrows}, {Campana}, {Gehrels}, {Guidorzi}, {Krimm}, {La Parola}, {Liang},
  {Mineo}, {Moretti}, {Page}, {Romano}, {Tagliaferri}, {Zhang}, {Page}, \&
  {Schady}}]{Troja2007}
{Troja} E. {et~al.}, 2007, \apj, 665, 599

\bibitem[{{Uhm}(2011)}]{Uhm2011}
{Uhm} Z.~L., 2011, \apj, 733, 86

\bibitem[{{Uhm} \& {Zhang}(2014)}]{Uhm2014}
{Uhm} Z.~L., {Zhang} B., 2014, \apj, 780, 82

\bibitem[{{Uhm} {et~al}\mbox{.}(2012){Uhm}, {Zhang}, {Hasco{\"e}t}, {Daigne},
  {Mochkovitch}, \& {Park}}]{Uhm2012}
{Uhm} Z.~L., {Zhang} B., {Hasco{\"e}t} R., {Daigne} F., {Mochkovitch} R.,
  {Park} I.~H., 2012, \apj, 761, 147

\bibitem[{{Usov}(1992)}]{Usov1992}
{Usov} V.~V., 1992, \nat, 357, 472

\bibitem[{{Van Eerten}(2013)}]{vanEerten2013proceedings}
{Van Eerten} H., 2013, 7th Huntsville Gamma-Ray Burst Symposium, GRB 2013:
  paper 24 in eConf Proceedings C1304143

\bibitem[{{van Eerten}(2014)}]{vanEerten2014correlations}
{van Eerten} H., 2014, Submitted. ArXiv e-prints: 1404.0283

\bibitem[{{Van Eerten} \& {MacFadyen}(2013)}]{vanEerten2013boostedcurves}
{Van Eerten} H., {MacFadyen} A., 2013, \apj, 767, 141

\bibitem[{{Van Eerten}, {van der Horst} \& {MacFadyen}(2012){Van Eerten}, {van
  der Horst}, \& {MacFadyen}}]{vanEerten2012boxfit}
{Van Eerten} H., {van der Horst} A., {MacFadyen} A., 2012, \apj, 749, 44

\bibitem[{{Van Eerten}, {Zhang} \& {MacFadyen}(2010){Van Eerten}, {Zhang}, \&
  {MacFadyen}}]{vanEerten2010offaxis}
{Van Eerten} H., {Zhang} W., {MacFadyen} A., 2010, \apj, 722, 235

\bibitem[{{Van Eerten} {et~al}\mbox{.}(2010){Van Eerten}, {Leventis},
  {Meliani}, {Wijers}, \& {Keppens}}]{vanEerten2010transrelativistic}
{Van Eerten} H.~J., {Leventis} K., {Meliani} Z., {Wijers} R.~A.~M.~J.,
  {Keppens} R., 2010, \mnras, 403, 300

\bibitem[{{Van Eerten} \& {MacFadyen}(2012)}]{vanEerten2012scalings}
{Van Eerten} H.~J., {MacFadyen} A.~I., 2012, \apjl, 747, L30

\bibitem[{{Van Eerten} \& {Wijers}(2009)}]{vanEerten2009}
{Van Eerten} H.~J., {Wijers} R.~A.~M.~J., 2009, \mnras, 394, 2164

\bibitem[{{Wijers} \& {Galama}(1999)}]{Wijers1999}
{Wijers} R.~A.~M.~J., {Galama} T.~J., 1999, \apj, 523, 177

\bibitem[{{Wijers}, {Rees} \& {Meszaros}(1997){Wijers}, {Rees}, \&
  {Meszaros}}]{Wijers1997}
{Wijers} R.~A.~M.~J., {Rees} M.~J., {Meszaros} P., 1997, \mnras, 288, L51

\bibitem[{{Woosley}(1993)}]{Woosley1993}
{Woosley} S.~E., 1993, \apj, 405, 273

\bibitem[{{Wu} {et~al}\mbox{.}(2003){Wu}, {Dai}, {Huang}, \& {Lu}}]{Wu2003}
{Wu} X.~F., {Dai} Z.~G., {Huang} Y.~F., {Lu} T., 2003, \mnras, 342, 1131

\bibitem[{{Yi}, {Wu} \& {Dai}(2013){Yi}, {Wu}, \& {Dai}}]{Yi2013}
{Yi} S.-X., {Wu} X.-F., {Dai} Z.-G., 2013, \apj, 776, 120

\bibitem[{{Zhang} {et~al}\mbox{.}(2006){Zhang}, {Fan}, {Dyks}, {Kobayashi},
  {M{\'e}sz{\'a}ros}, {Burrows}, {Nousek}, \& {Gehrels}}]{ZhangBing2006}
{Zhang} B., {Fan} Y.~Z., {Dyks} J., {Kobayashi} S., {M{\'e}sz{\'a}ros} P.,
  {Burrows} D.~N., {Nousek} J.~A., {Gehrels} N., 2006, \apj, 642, 354

\bibitem[{{Zhang} \& {Kobayashi}(2005)}]{ZhangBing2005}
{Zhang} B., {Kobayashi} S., 2005, \apj, 628, 315

\bibitem[{{Zhang} \& {M{\'e}sz{\'a}ros}(2001)}]{ZhangMeszaros2001}
{Zhang} B., {M{\'e}sz{\'a}ros} P., 2001, \apjl, 552, L35

\bibitem[{{Zhang} \& {MacFadyen}(2009)}]{Zhang2009}
{Zhang} W., {MacFadyen} A., 2009, \apj, 698, 1261

\bibitem[{{Zhang} \& {MacFadyen}(2006)}]{ZhangWeiqun2006}
{Zhang} W., {MacFadyen} A.~I., 2006, \apjs, 164, 255

\bibitem[{{Zou} \& {Piran}(2010)}]{ZouPiran2010}
{Zou} Y.-C., {Piran} T., 2010, \mnras, 402, 1854

\bibitem[{{Zou}, {Wu} \& {Dai}(2005){Zou}, {Wu}, \& {Dai}}]{Zou2005}
{Zou} Y.~C., {Wu} X.~F., {Dai} Z.~G., 2005, \mnras, 363, 93

\end{thebibliography}

\appendix

\onecolumn

\section{Additional symbols self-similar solution}
\label{additional_symbols_appendix}

For completenes, I define in this appendix some symbols used in expressing the self-similar solution in section \ref{self-similar_blast_wave_subsection}. Straight out of BM76 we have
\begin{equation}
A \equiv \frac{x^2 + 2 \alpha_1 x - 8 \beta_1}{1 + 2 \alpha_1 - 8 \beta_1}, \; \;
B \equiv \frac{(x + \alpha_1 + \gamma_1)(1 + \alpha_1 - \gamma_1)}{(x + \alpha_1 - \gamma_1)(1 + \alpha_1 + \gamma_1)}.
\end{equation}
\begin{equation}
\alpha_1 = \frac{m - 3k+12}{2}, \; \; \beta_1 = \frac{m+1}{2}, \; \; \gamma_1 = \sqrt{\alpha_1^2 + 8 \beta_1},
\end{equation}
\begin{equation}
 \alpha_2 = \frac{-(m+k-4)}{2}, \; \; \beta_2 = \frac{(m-3k)(m+k) + 8(3m+4k-8)}{4}, \; \; \gamma_2 = \frac{-(m^2 + 4mk + 3k^2 -13m -19k + 24)}{2(m+3k-12)},
\end{equation}
\begin{equation}
\mu_1 = \frac{2(7m^2+34mk-118m+15k^2 -82k + 96) + (m - 3k + 12)(m^2 + 4mk +3k^2 -13m - 19k + 24)}{4(m+3k-12)},
\end{equation}
\begin{equation}
\mu_2 = \frac{m-k}{m+3k-12}.
\end{equation}
In addition to the analytical solution eq. \ref{h_solution_equation} for the density profile, we have for the pressure profile $f$ and Lorentz factor profile $g$:
\begin{equation}
f = A(x)^{-\alpha_2} \times B(x)^{\beta_2 / \gamma_1}, \; \; g = A(x)^{\frac{1}{2} + \beta_1} \times B(x)^{[ \alpha_1 (\beta_1 - 1/2) + 8 \beta_1] / \gamma_1}.
\end{equation}

\section{Description of numerical code and settings}
\label{code_appendix}

All simulations were performed using the parallel adaptive-mesh refinement (AMR) relativistic hydrodynamics code \textsc{ram} \citep{ZhangWeiqun2006, Zhang2009}. \textsc{Ram} makes use of the \textsc{paramesh} amr tools \citep{MacNeice2000} from \textsc{flash} 2.3 \citep{Fryxell2000}. The code allows for various solvers and coordinate systems. For this study, spherical coordinates are used. The RHD equations are solved using a piecewise linear method (a practical approach in the presence of strong shocks). A relativistic equation of state is used with adiabatic index $\Gamma_{ad} = 4/3$ (all relevant parts of the outflow are ultra-relativistic, the BM solution also makes use of this).

The same grid size is used for each simulation. The lower and upper radial boundaries are at $2.998 \times 10^{13}$ cm ($10^3$ ls) and $1.499 \times 10^{18}$ cm ($5 \times 10^7$ ls) respectively. This grid is divided into 2 base level blocks of 8 cells each and further dynamically subdivided up to 21 levels of refinement (aside from lower refinement consistency checks), where and when needed. As a result, the effective resolution $\delta$cell is $8.934 \times 10^{10}$ cm ($2.980$ ls). For comparison, the width $\Delta R$ of the shell at $t = T_{last}$ is $1.804 \times 10^{14}$ cm ($6.016 \times 10^3$ ls) for the typical $k = 2$ case and $1.887 \times 10^{14}$ cm ($6.300 \times 10^{3}$ ls) for the typical $k = 0$ case. The RS-CD-FS system is therefore well resolved at all relevant times. The refinement level of the RS-CD-FS system is manually kept at peak level, while regions at small radii, compared to either the reverse shock radius or the current radius of the last injected energy, are automatically derefined.

The grid is set up with a cold fluid with the appropriate profile (i.e. $k = 0, 2$) and the energy is injected via a boundary condition on the lower boundary, given by eqs. \ref{freely_expanding_flow_equations}. After a time $T_{stop}$ (corresponding to an energy injection duration $T_{in}$ and accounting for the fact that the last injected energy has to be beyond the inner boundary), the luminosity is decreased exponentially according to $L_{drop} = L (\exp[- (t - T_{stop}) / (100$ $T_{stop})] + 10^{-5})$. The mass inflow $L_M$ is equally decreased, while $\eta$ is kept fixed. As long as the energy injection is decreased sufficiently fast, the exact post-injection evolution of $L$, $L_M$ and $\eta$ will not impact the outcome and the current set-up is chosen for numerical reasons.

\section{Heuristic flux fit functions}
\label{heuristic_flux_appendix}

\begin{table}
\begin{center}
\begin{tabular}{r|l|l|l|l|l|l|}
\hline
 & $F_{peak,FS}$ & $\nu_{m,FS}$ & $\nu_{c,FS}$ & $F_{peak,RS}$ & $\nu_{m,RS}$ & $\nu_{c, RS}$ \\
 pre-factor & & $[(2-p)/(1-p)]^2$ & & & $[(2-p)/(1-p)]^2$ & \\
 0 & $7.968$ & $-4.646$ & $-4.136$ & $3.691$ & $0.538$ & $-1.668$ \\
 $p$ & $0.381$ & $0.063$ & $0.122$ & $0.393$ & $0.110$ & $-0.118$ \\
 $k$ & $-3.484$ & $1.522$ & $5.512$ & $-1.173$ & $-2.057$ & $3.984$ \\
 $q$ & $3.577$ & $2.339$ & $-1.565$ & $6.359$ & $-0.065$ & $-2.387$ \\
 $kq$ & $-1.200$ & $-0.696$ & $1.055$ & $-1.751$ & $-0.525$ & $1.525$ \\
 $k^2$ & $0.126$ & $-0.325$ & $-0.443$ & $-0.006$ & $-0.106$ & $-0.234$ \\
 $q^2$ & $0.135$ & $0.101$ & $0.158$ & $-0.720$ & $0.043$ & $0.348$ \\
 $pk$ & $-0.123$ & $-0.099$ & $-0.185$ & $-0.142$ & $-0.158$ & $0.196$ \\
 $pq$ & $-0.029$ & $-0.085$ & $-0.028$ & $-0.030$ & $-0.089$ & $0.074$ \\
 $k^2q$ & $0.004$ & $-0.010$ & $-0.022$ & $-0.008$ & $-0.007$ & $-0.008$ \\
 $kq^2$ & $-0.133$ & $0.045$ & $-0.003$ & $0.150$ & $-0.010$ & $-0.103$ \\
 $pqk$ & $0.016$ & $0.039$ & $0.108$ & $0.024$ & $0.071$ & $-0.043$ \\
 $pk^2$ & $0.022$ & $0.064$ & $0.044$ & $0.026$ & $0.077$ & $-0.048$ \\
 $pq^2$ & $-0.000$ & $0.023$ & $-0.081$ & $0.006$ & $0.017$ & $-0.009$ \\
\hline
\end{tabular}
\caption{Coefficients to heuristic functions for approximating the flux level. See main appendix text for instructions on usage.}
\label{injection_polynomial_terms_table}
\end{center}
\end{table}

The equations in tables \ref{characteristics_scalings_table} and \ref{flux_scalings_table} show how the flux depends on the model parameters, but do not give the absolute flux levels. An approximate heuristic function to the flux levels is provided in table \ref{injection_polynomial_terms_table}, calculated from the asymptotic limits for cases where $\nu_m \ll \nu_c$. The lines in this table specify the coefficients $\alpha_i$ of the function
\begin{eqnarray}
C(p,q,k) & \equiv & \alpha_0 + \alpha_p p + \alpha_k k + \alpha_q q + \alpha_{kq} kq + \alpha_{kk} k^2 + \alpha_{qq} q^2 + \nonumber \\
 & & \alpha_{pk} pk + \alpha_{pq} p q + \alpha_{kkq} k^2 q + \alpha_{kqq} k q^2 + \alpha_{pqk} pqk + \alpha_{pkk} pk^2 + \alpha_{pqq} pq^2.
\end{eqnarray}
For $F_{peak,FS}$, this yields $C(p,q,k) = 8.080 + 0.338 p - 3.798k \ldots$, etc.

The values of the characteristic quantities are then given by
\begin{eqnarray}
F_{peak,FS} & = & \textrm{pre-factor } \times 10^{C(p,q,k)/(4-k)} \frac{\left( 1+z \right)^{\frac{k-8}{2(k-4)}}}{d_{L,28}^2} \xi_{N,0} \epsilon_{B,-2}^{\frac{1}{2}} \left(n_{ref,0} R_{ref,17}^k \right)^{\frac{-2}{k-4}} \times \nonumber \\
  & & \left( E_{iso,53} T_{in,4}^{-(1+q)} \right)^{\frac{-(8-3k)}{2(k-4)}} t_{obs,0}^{\frac{4k - 8 -8q + 3kq}{2(k-4)}} \textrm{mJy}, \nonumber \\
\nu_{m,FS,15} & = & \textrm{pre-factor } \times 10^{C(p,q,k)/(4-k)} (1+z)^{\frac{1}{2}} \xi_{N,0}^{-2} \epsilon_{e,-1}^2 \epsilon_{B,-2}^{\frac{1}{2}} \left( E_{iso,53} T_{in,4}^{-(1+q)} \right)^{\frac{1}{2}} t_{obs,0}^{\frac{q-2}{2}},
\label{poly_equations}
\end{eqnarray}
et cetera. Here I use the notation $X_{-2} \equiv X \times 10^{-2}$ etc., in cgs units (except for observer time in days), in order to express the characteristic quantities and table entries relative to a given base level. Using the tabulated values will yield the characteristic frequencies in units of $10^{15}$ Hz (i.e. $\nu_{15}$) and peak fluxes in mJy. The base level values of the model parameters are $d_L = 10^{28}$ cm, $\eta = 1000$, $\xi = 1$, $\epsilon_e = 10^{-1}$, $\epsilon_B = 10^{-2}$, $n_{ref} = 1$ cm$^{-3}$, $R_{ref} = 10^{17}$ cm, $E_{iso} = 10^{53}$ erg, $T_{in} = 10^4$ s, $t_{obs} = 1$ day, hence the notation $E_{iso,53}$, $\xi_0$, etc. in eqs. \ref{poly_equations} above. Note that these base line values are not always identical to the \emph{typical} values first used in section \ref{typical_values_section}. 

The heuristic functions have no physical meaning and are merely the combination of a straightforward polynomial divided by the $(4-k)$ factor common to many other terms in the expressions for the characteristic quantities. Their dependence on $p$ is not strong (and only terms up to first order in $p$ are therefore used). In a homogeneous shell model the dependence on $p$ drops out of the characteristic quantity equations completely (see e.g. \citealt{Leventis2014}). Nor does $p$ occur in the exponents of the model parameter dependencies in the characteristic equations for a non-homogeneous profile (see \ref{characteristics_scalings_table}). When not ignored in the pre-factors too (e.g. in \citealt{vanEerten2012scalings, vanEerten2013boostedcurves}), the residual $p$-dependency in the pre-factors is the result of the measured characteristic quantities being the weighed average of their values at all simultaneously observed positions in the fluid, with the weighing function being the $p$-dependent local synchrotron spectrum.

The heuristic functions have been determined by minimizing their differences to the simulation-derived values at all 27 permutations of $k = (0, 1, 2)$, $q = (-0.5, 0, 1)$ and $p = (2.1, 2.5, 3)$. With the values from table \ref{injection_polynomial_terms_table}, the differences always remain at or below $3\%$.

\begin{table}
\begin{center}
\begin{tabular}{r|l|l|l|l|l|l|l|}
\hline
 & pre-factor & $p$ & $k$ & $k^2$ & $pk$ & $pk^2$ \\
 $F_{peak, I}$ & 1 & 4.340 & 0.508 & -2.443 & 0.133 & -0.149 \\
 $\nu_{m, I}$ & $[(2-p)/(1-p)]^2$ & -7.299 & 0.456 & 2.192 & -0.173 & -0.181 \\
 $\nu_{c, I}$ & 1 & -2.659 & -0.037 & 4.829 & -0.464 & -0.338 \\
\hline
\end{tabular}
\caption{Same as table \ref{injection_polynomial_terms_table}, now for impulsive injection.}
\label{impulsive_polynomial_terms_table}
\end{center}
\end{table}

For comparison, I also provide in table \ref{impulsive_polynomial_terms_table} heuristic functions describing the flux level for impulsive energy injection blast waves. These can be combined with flux equations in the various regimes, constructed from table \ref{characteristics_scalings_table} or taken from \cite{vanEerten2009}. Note that in \cite{vanEerten2009}, a different approach was taken to the synchrotron function, leading to differences in flux levels between that study and table \ref{impulsive_polynomial_terms_table}.

\bsp

\label{lastpage}

\end{document}